\documentstyle[seceq,preprint,mbf]{ptptex}
\def\slash#1{\ooalign{\hfil/\hfil\crcr$#1$}}

\def\XZ{Z}
\def\XP{P}
\def\XQ{Q}
\def\XM{M}
\def\XD{D}
\def\XS{S}
\def\XK{K}
\def\XU{U}
\def\XX{X}
\def\XG{G}
\def\XA{{\mbf A}}
\def\XV{{\mbf V}}
\def\XH{{\mbf H}}
\def\XL{{\mbf L}}

\def\dt{\!\cdot\!}
\def\nn{\nonumber\\}

\def\calA{{\cal A}}
\def\calD{{\cal D}}
\def\calN{{\cal N}}
\def\hatD{{\hat\calD}}

\def\slashD{{\hat{\slash\calD}}}

\def\hatR{{\hat R}}

\def\calF{{\cal F}}

\def\calR{{\cal R}}

\def\myfrac#1#2{\hbox{\large ${#1\over #2}$}}
\def\T{{\rm T}}

\def\pr#1{{#1'}}

\def\6#1{{\underline{#1}}}
 \def\m6#1{{\underline{#1}\,}}

\newdimen\Tdim
\def\ispan{{\setbox0=\hbox{i}%
\Tdim\ht0\advance\Tdim\dp0\rule[-\dp0]{0pt}{\Tdim}}}
\def\jspan{{\setbox0=\hbox{j}%
\Tdim\ht0\advance\Tdim\dp0\rule[-\dp0]{0pt}{\Tdim}}}
\def\Tspan#1{{\setbox0=\hbox{#1}%
\Tdim\ht0\advance\Tdim\dp0\advance\Tdim.55ex\rule[-\dp0]{0pt}{\Tdim}\box0}}
\newcommand{\beq}{\begin{equation}}
\newcommand{\eeq}{\end{equation}}
\newcommand{\beqa}{\begin{eqnarray}}
\newcommand{\eeqa}{\end{eqnarray}}
\newcommand{\eqn}[1]{&#1&}
\newcommand{\bear}{\begin{array}}
\newcommand{\eear}{\end{array}}
\newcommand{\CR}{\nonumber \\ }

\newcommand{\cD}{{\cal D}}

\newcommand{\cO}{{\cal O}}

\newcommand{\alp}{\alpha}

\newcommand{\gam}{\gamma}
\newcommand{\Gam}{\Gamma}
\newcommand{\eps}{\epsilon}
\newcommand{\lam}{\lambda}

\newcommand{\Ome}{\Omega}
\newcommand{\veps}{\varepsilon}

\newcommand{\vphi}{\varphi}

\newcommand{\mn}{\mu\nu}
\newcommand{\rs}{\rho\sigma}

\newcommand{\undb}{\underline{b}}

\newcommand{\unde}{\underline{e}}
\newcommand{\undf}{\underline{f}}

\newcommand{\uA}{\underline{A}}
\newcommand{\uC}{\underline{C}}
\newcommand{\uD}{\underline{D}}
\newcommand{\uE}{\underline{E}}
\newcommand{\uL}{\underline{L}}
\newcommand{\uM}{\underline{M}}
\newcommand{\uN}{\underline{N}}

\newcommand{\uT}{\underline{T}}
\newcommand{\uV}{\underline{V}}
\newcommand{\uW}{\underline{W}}
\newcommand{\uY}{\underline{Y}}

\newcommand{\bareta}{\overline{\eta}}

\newcommand{\bveps}{\overline{\varepsilon}}
\newcommand{\undvphi}{\underline{\vphi}}
\newcommand{\blam}{\overline{\lambda}}\newcommand{\undlam}{\underline{\lam}}

\newcommand{\uPhi}{\underline{\Phi}}
\newcommand{\undphi}{\underline{\phi}}
\newcommand{\bpsi}{\overline{\psi}}\newcommand{\undpsi}{\underline{\psi}}

\newcommand{\undchi}{\underline{\chi}}

\newcommand{\undome}{\underline{\omega}}
\newcommand{\bOme}{\overline{\Omega}}\newcommand{\uOme}{\underline{\Omega}}
\newcommand{\undzeta}{\underline{\zeta}}

\newcommand{\hOme}{\hat{\Omega}}
\newcommand{\hF}{\hat{F}}

\newcommand{\hR}{\hat{R}}\newcommand{\uhR}{\underline{\hR}}

\newcommand{\chD}{\hat{\cal D}}

\newcommand{\nchD}{\chD\hspace{-0.60em}/\,}

\preprintnumber[3cm]{
KUNS-1716
\\ hep-th/0104130}

\title{
Superconformal Tensor Calculus\\ in Five Dimensions 
}

\author{
Tomoyuki {\sc Fujita}\footnote{E-mail:
fujita@gauge.scphys.kyoto-u.ac.jp}
and Keisuke {\sc Ohashi}\footnote{E-mail:
keisuke@gauge.scphys.kyoto-u.ac.jp}
}

\inst{
Department of Physics, Kyoto University, Kyoto 606-8502, Japan
}

\recdate{
\today
}

\abst{
We present a full superconformal tensor calculus in five spacetime
dimensions in which the Weyl multiplet has 32 Bose plus 32 Fermi
degrees of freedom. 
It is derived using dimensional reduction from the 6D
superconformal tensor calculus.
We present two types of 32+32 Weyl multiplets,
a vector multiplet, linear multiplet,
hypermultiplet and nonlinear multiplet.
Their superconformal transformation laws and
the embedding and invariant action formulas are given.}

\begin{document}
\maketitle

\section{Introduction}
Five-dimensional supergravity,
which was once extensively studied,\cite{Cre} has recently
received much attention again.\cite{Gun-Zag,Cer-Dal,Nis-Raj,Ber-Her}
This renewed interest is due partly to the study of
the AdS/CFT correspondence conjecture.
This conjecture suggests that
gauged supergravity in a background geometry
AdS$_5\times H$ ($H=S^5$ in the original example of Ref.~\citen{Mal})
is related to superconformal field theory
in a four-dimensional Minkowski space on the
boundary of AdS$_5$.

Another line of investigation that has motivated
study of five-dimensional supergravity
is the search for supersymmetric brane-worlds scenarios.
In particular, from both phenomenological and theoretical
viewpoints,
it is interesting to supersymmetrize
Randall-Sundrum scenarios.\cite{RS1,RS2}
The simplest candidate for the supersymmetric Randall-Sundrum
two branes model, namely RS1,\cite{RS1}
is five-dimensional supergravity compactified on
the $S^1/Z^2$ orbifold.
In the five-dimensional bulk, there exists
a minimal or nonminimal
supergravity multiplet \cite{Nis-Raj}
that contains a graviton, gravitino and graviphoton.
This multiplet is trapped on the branes,
reduces to the four-dimensional minimal multiplet, and
couples to the four-dimensional matter multiplets,
e.g., the chiral and vector multiplets.
Further, we can couple this multiplet
to various matter multiplets in the bulk,
for example the vector, hyper and tensor multiplets.
In order to work with these models,
off-shell formalisms \cite{Kug-Oha1,Kug-Oha2,Zuk12,Zuk3}
rather than on-shell formalisms
facilitate the analysis, because with these,
we need not change the transformation laws of the supersymmetry,
whichever couplings are considered.
Still, it is laborious
to study a large class of such models systematically.
However the gauge equivalence method 
using superconformal tensor calculus makes this task easy.
In this formulation, to construct different off-shell formulations,
we have only to add a compensating multiplet
to the Weyl multiplet, so
we can treat all of the above mentioned couplings in a common framework.
However, unfortunately, five-dimensional conformal supergravity
has not yet been studied.

Standard conformal supergravity can be
described on the basis of superconformal algebra.
Superconformal algebra exists only in six or
fewer dimensions,\cite{Nah}
and its gauge theory has been constructed
in the case of 16 or fewer
supercharges,\cite{n1d41,n2d41,n4d4,Ber-Sez-Pro,n4d6}
except for the case of $N=2, d=5$ theory, in which we are interested.
(Five-dimensional conformal supergravity
that is not based on superconformal algebra
was constructed through dimensional reduction from
ten-dimensional conformal supergravity.\cite{Ber-deR-deW})
In this paper we fill the gap in the literature
by constructing $N=2, d=5$ superconformal tensor calculus
in a complete form.

In Ref.~\citen{Kug-Oha1},
5D tensor calculus was derived from the known
6D superconformal tensor calculus \cite{Ber-Sez-Pro}
using the method of dimensional reduction.
However, unfortunately,
some of the superconformal symmetries ($S$ and $K$)
are gauge-fixed in the process of the reduction.
The dimensional reduction is in principle
straightforward and hence more convenient than
the conventional trial-and-error method to find the
multiplet members and their transformation laws.
Therefore, here we follow essentially the
dimensional reduction used in Ref.~\citen{Kug-Oha1}
to find the 5D superconformal tensor calculus
from the 6D one.
We keep all the 5D superconformal gauge symmetries
unfixed in the reduction process.
The Weyl multiplet obtained from a simple reduction
contains 40 bose and 40 fermi degrees of freedom.
However, it turns out that this $40+40$ multiplet
splits into two irreducible pieces,
a $32+32$ minimal Weyl multiplet
and an $8+8$ `central charge' vector multiplet
(which contains a `dilaton' $e_z{}^5$
and a graviphoton $\propto e_{\mu}{}^5$
as its members).
This splitting is performed by inspecting and comparing
the transformation laws of both the Weyl and vector multiplets.
It contains a process
that is somewhat trial-and-error in nature,
but can be carried out relatively easily.
Once this minimal Weyl multiplet is found,
the other processes of finding matter multiplets and
other formulas, like invariant action formulas,
proceed straightforwardly and are very similar to
those in the previous Poincar\'{e}
supergravity case.\cite{Kug-Oha1}

For the reader's convenience,
we give the details of the dimensional reduction procedure
in Appendix B
and present the resultant transformation law of the
minimal $32+32$ Weyl multiplet in \S 2.
The transformation rules of the matter multiplets
are given in \S 3;
the multiplets we discuss are the vector (Yang-Mills)
multiplet, linear multiplet, hypermultiplet
and nonlinear multiplet.
In \S 4, we present some embedding formulas of
multiplets into multiplet and invariant action formulas.
In \S 5, we present another $32+32$ Weyl multiplet
that corresponds to the Nisino-Rajpoot version of
Poincar\'{e} supergravity.
This multiplet is expected to appear by dimensional reduction
from the `second version' of 6D Weyl multiplet containing a tensor
$B_{\mn}$ and by separating the $8+8$ vector multiplet.
Here, however, we construct it directly in 5D by
imposing a constraint on a set of vector multiplets.
Section 6 is devoted to summary and discussion.
The notation and some useful formulas are
presented in Appendix A.
In Appendix D, we explain the relation between conformal
supergravity constructed in this paper and
Poincar\'e supergravity worked out in Ref.~\citen{Kug-Oha1}.

\section{Weyl multiplet}
The superconformal algebra in five dimensions is $F^2(4)$.\cite{Nah}
Its Bose
sector is $SO(2,5)\oplus SU(2)$. 
The generators of this algebra $\XX_{\bar A}$ are
\begin{equation}
\XX_{\bar A}=\quad \XP_a,\quad \XQ_i,\quad \XM_{ab},\quad \XD,\quad 
\XU_{ij},\quad \XS^i,\quad \XK_a,
\end{equation}
where $a,b,\ldots$ are Lorenz indices, $i,j,\ldots(=1,2)$ are $SU(2)$
indices, and $\XQ_i$ and $\XS_i$ have spinor indices implicitly.  
$\XP_a$ and $\XM_{ab}$ are the usual Poincar\'e generators, $\XD$ is the
dilatation, $\XU_{ij}$ is the $SU(2)$ generator,
 $\XK_a$ represents the special conformal boosts, $\XQ_i$ represents 
the $N=2$ supersymmetry, and $\XS_i$ represents the conformal
supersymmetry. The gauge fields $h^A_\mu $ corresponding 
to these generators are 
\begin{equation}
h_\mu ^{\bar A}=\quad e_\mu {}^a,\quad \psi _\mu ^i,\quad \omega _\mu {}^{ab},
\quad b_\mu ,\quad V_\mu ^{ij},\quad \phi _\mu ^i,\quad f_\mu {}^a,
\end{equation}
respectively, where $\mu ,\nu ,\ldots$ are the world vector indices and $\psi _\mu ^i, \phi _\mu ^i$ are $SU(2)$-Majorana
spinors. (All spinors satisfy the $SU(2)$-Majorana condition in this
calculus.\footnote{Only a spinor of the hypermultiplet $\zeta _\alpha $ is not
such a spinor, but a $USp(2n)$-Majorana spinor.})
In the text we omit explicit expression of the covariant derivative
$\hatD_\mu $ and the covariant curvature $\hatR_{\mu \nu }{}^A$ (field
strength $\hat F_{\mu \nu }{}^A$). In our calculus,
the definitions of these are given as follows:\cite{Kug-Oha1}  
\begin{equation}
\hatD_\mu \Phi \equiv \partial _\mu \Phi -h_\mu ^AX_A\Phi ,
\end{equation}
\begin{equation}
\hatR_{\mu \nu }{}^{\bar A}=e_\mu {}^be_\nu ^af_{ab}{}^{\bar A}
=2\partial _{[\mu }h_{\nu ]}{}^{\bar A}
-h_\mu ^{\bar C}h_\nu ^{\bar B}\pr f_{\bar B\bar C}{}^{\bar A}.\label{a}
\end{equation}
Here, $X_A$ denotes the transformation
operators other than $P_a$, and $f_{\bar A\bar B}{}^{\bar C}$ is a
`structure function', defined by $[X_{\bar A},\,X_{\bar B}\}=
f_{\bar A\bar B}{}^{\bar C}X_{\bar C}$,
which in general depends on the fields. The prime on the structure
function in (\ref{a}) indicates that the $[P_a,\,P_b]$ commutator part,
$f_{ab}{}^{\bar A}$, is excluded from the sum. 
Note that this structure function can be read from the transformation
laws of the gauge fields: 
$\delta (\varepsilon )h_\mu ^{\bar A}=\delta _{\bar B}(\varepsilon ^{\bar B})h_\mu ^{\bar A}
=\partial _\mu \varepsilon ^{\bar A}+\varepsilon ^{\bar C}h_\mu ^{\bar B}f_{\bar B\bar C}{}^{\bar A}$.
\subsection{Constraints and the unsubstantial gauge fields}
In the superconformal theories in 4D and 6D,
\cite{n1d41,n2d41,n4d4,Ber-Sez-Pro,n4d6}  the conventional constraints on
the superconformal curvatures are imposed to lift the tangent-space
transformation $P_a$ to the general coordinate transformation of a Weyl multiplet. 
These constraints are the usual torsion-less condition,
\begin{equation}
\hatR_{ab}{}^c(P)=0,\label{eq:constr.P}
\end{equation}
and two conditions on $Q$ and $M$ curvatures
of the following types:
\begin{equation}
\gamma ^b\hatR_{ab}(Q)=0,\qquad \hatR_{ac}{}^{cb}(M)=0.\label{eq:constr.SK}
\end{equation}
The spin-connection $\omega _\mu {}^{ab}$ becomes a dependent field by the
constraint (\ref{eq:constr.P}), and the $\XS_i$ and $\XK_a$
gauge fields, $\phi_\mu^i$ and $f_\mu {}^a$,
also become dependent through the constraints
(\ref{eq:constr.SK}). 
To this point, it has been the conventional understanding that imposing these
curvature constraints is unavoidable for the purpose
of obtaining a meaningful local
superconformal algebra. However, it is actually possible to
avoid imposing the constraints explicitly, but we can obtain an
equivalent superconformal algebra. This fact is not familiar, so for a 
better understanding we illustrate this approach with the
transformation laws of the well-known $N=1,d=4$ Weyl multiplet in Appendix 
C. We now explain how this is possible 
by considering an example. In 5D, the covariant derivative of the spinor 
$\Omega ^i$ of a vector multiplet and the $\XQ$ curvature contain the $\XS_i$ 
gauge field $\phi _\mu ^i$ in the form
  \begin{eqnarray}
\left.\slashD\Omega ^i\right|_{\phi \rm -term}=M\gamma ^a\phi _a^i,
\left.\qquad \gamma \dt \hatR^i(Q)\right|_{\phi \rm -term}=8\gamma ^a\phi _a^i.
\end{eqnarray}
However, in fact, in the supersymmetry transformation $\delta Y^{ij}$
of the auxiliary field $Y^{ij}$ of the vector multiplet, only the combination
\begin{equation}
{\cal C}\equiv 2\slashD\Omega ^i-\myfrac14\gamma \dt \hatR^i(Q)M 
\end{equation}
appears, and the gauge field $\phi _\mu ^i$ is
actually canceled in these two
terms. Since this combination contains no $\phi _\mu ^i$, we can set $\phi _\mu 
^i$ equal to anything. For instance, we can set $\phi _\mu ^i=\phi _\mu ^{i\,\rm
sol}$, the solution $\phi _\mu $ to the conventional constraint 
$\gamma ^a\hatR_{ab}{}^i(Q)=0$. Then the $\gamma \dt\hatR^i(Q)$ term
vanishes, and the combination clearly reduces to 
\begin{eqnarray}
{\cal C}=\left.2\slashD\Omega ^i\right|_{\phi _\mu \rightarrow \phi _\mu ^{\rm sol}},
\end{eqnarray}
reproducing the result of the conventional approach. The
virtue of our approach is, however, that it is independent of the form 
of the constraints. If the constraints are changed into $\gamma 
^a\hatR_{ab}{}^i(Q)=\gamma _b\pr \chi ^i$,
with a certain spinor $\pr \chi ^i$, then the combination takes an
apparently different form,
\begin{equation}
{\cal C}=\left.2\slashD\Omega ^i\right|_{\phi _\mu \rightarrow \pr \phi _\mu ^{\rm sol}}
+\myfrac54\pr \chi ^iM.
\end{equation}
Everywhere in this calculus, in the transformation laws,
the algebra, the embedding formulas, the action formulas, and so on, 
such cancellations occur, 
so the gauge fields $\phi _\mu ^i$ and $f_\mu {}^a$ actually 
disappear completely. 
 
 In this 5D calculus, we adopt the usual torsion-less condition
(\ref{eq:constr.P}), but we do not impose constraints on the
 $\XQ_i$ and $\XM_{ab}$ curvatures, because no such constraints
significantly simplify the 5D calculus, and the formulation with no
constraint is convenient to 
reduce Poincar\'e supergravity calculus from this conformal one.
 We comment on these reductions in Appendix D. 
To make the expressions brief,
we define the covariant quantities $\phi _a^i(Q),f_a{}^b(M),K_{ab}(Q)$
as
\begin{eqnarray}
\phi ^i_a(Q)&\equiv &\myfrac13\gamma ^b\hatR^i_{ab}(Q)
-\myfrac1{24}\gamma _a\gamma \dt \hatR^i(Q),\nn
f_{ab}(M)&\equiv &-\myfrac16\hatR_{ab}(M)+\myfrac1{48}\eta _{ab}\hatR(M),\nn
K_{ab}(Q)&\equiv &\hatR_{ab}(Q)+2\gamma _{[a}\phi _{b]}(Q)\nn
&=&\hatR_{ab}(Q)+\myfrac23\gamma _{[a}\gamma ^c\hatR_{b]c}
-\myfrac1{12}\gamma _{ab}\gamma \dt  \hatR(Q),
\end{eqnarray}
where, $\hatR_{ab}(M)\equiv \hatR_{ac}{}^{cb}(M), \hatR(M)\equiv \hatR_a{}^a(M)$.
These quantities are defined in such a way that they contain the $S$
and $K$ gauge fields in the simple forms
\begin{equation}
\left.\phi _a^i(Q)\right|_{\phi ,f}=\phi _a^i,\quad 
\left.f_{ab}(M)\right|_{\phi ,f}=f_{ab},
\left.\quad K_{ab}(Q)\right|_{\phi ,f}=0,
\end{equation}
and  $K_{ab}(Q)$ satisfies
\begin{equation}
\gamma ^aK_{ab}(Q)=0.
\end{equation}
Since we impose the torsion-less constraint (\ref{eq:constr.P}) in 5D too, 
the spin-connection is a dependent field given by
\begin{eqnarray}
\omega _\mu {}^{ab}&=&\omega ^0_\mu {}^{ab}
+i(2\bar\psi _\mu \gamma ^{[a}\psi ^{b]}+\bar\psi ^a\gamma _\mu \psi ^b)-2e_\mu {}^{[a}b^{b]},\nn
\omega ^0_\mu {}^{ab}&\equiv &-2e^{\nu [a}\partial _{[\mu }e_{\nu ]}{}^{b]}
+e^{\rho [a}e^{b]\sigma }e_\mu {}^c\partial _\rho e_{\sigma c}.\label{eq:spincon}
\end{eqnarray}
Of course, it would also be possible
to avoid this torsion-less constraint in a
similar way, but here we follow the conventional procedure.  

\subsection{The transformation law and the superconformal algebra}
\begin{table}[tb]
\caption{Weyl multiplet in 5D.}
\label{table:5DWeyl}
\begin{center}
\begin{tabular}{ccccc} \hline \hline
    field      & type   & remarks & {\it SU}$(2)$ & Weyl-weight    \\ \hline 
\Tspan{$e_\mu{}^a$} &   boson    & f\"unfbein    & \bf{1}    &  $ -1$     \\  
$\psi^i_\mu$  &  fermion  & {\it SU}$(2)$-Majorana & \bf{2}
&$-\frac{1}{2}$ \\  
\Tspan{$b_\mu$} & boson &  real & \bf{1} & 0 \\
\Tspan{$V^{ij}_\mu$}    &  boson    & $V_\mu^{ij}=V_\mu^{ji}=(V_{\mu ij})^*$ 
& \bf{3}&0\\ 
$v_{ab}$&boson& real, antisymmetric &\bf{1}&1 \\
$ \chi^i$  &  fermion  & {\it SU}$(2)$-Majorana & \bf{2}    &$\frac{3}{2}$ \\  
$D$    &  boson    & real & \bf{1} & 2 \\ \hline
\multicolumn{5}{c}{dependent (unsubstantial) gauge fields} \\ \hline
$\omega_\mu{}^{ab}$ &   boson    & spin connection & \bf{1}    &   0\\
$\phi _\mu ^i$ & fermion & SU(2)-Majorana & \bf{2} & $\frac{1}{2}$\\
$f_\mu {}^a$ & boson & real & \bf{1} & 1\\
\hline 
\end{tabular}
\end{center}
\end{table}
The superconformal tensor calculus in 5D can be obtained from the
known one in 6D by carrying out a simple dimensional reduction.
However, the 
Weyl multiplet directly obtained this way contains $40+40$ degrees of 
freedom. Using the procedure explained in detail in Appendix B, we
can separate an $8+8$ component vector multiplet from it and obtain an 
irreducible Weyl multiplet which
consists of 32 Bose plus 32 Fermi fields,
\begin{eqnarray}
e_\mu {}^a,\quad \psi _\mu ^i,\quad V_\mu ^{ij},\quad b_\mu ,\quad v^{ab},\quad 
        \chi ^i,\quad D,
\end{eqnarray}
whose properties are summarized in Table \ref{table:5DWeyl}.
The full nonlinear $\XQ, \XS$ and $ \XK$ transformation laws of the Weyl
multiplet are given as follows. With
 $\delta \equiv \bar\varepsilon ^i\XQ_i+\bar\eta ^i\XS_i+\xi _K^a\XK_a\equiv \delta _Q(\varepsilon )+
\delta _S(\eta )+\delta _K(\xi _K^a)$,
\begin{eqnarray}
\delta e_\mu {}^a&=&-2i\bar\varepsilon \gamma ^a\psi _\mu ,\nn
\delta \psi _\mu ^i&=&{\cal D}_\mu \varepsilon ^i+\myfrac12 v^{ab}\gamma _{\mu ab}\varepsilon ^i-\gamma _\mu \eta ^i,\nn
\delta b_\mu &=&-2i\bar\varepsilon \phi _\mu +2i\bar\varepsilon \phi _\mu (Q)-2i\bar\eta \psi _\mu -2\xi _{K\mu },\nn
\delta \omega _\mu {}^{ab}&=&2i\bar\varepsilon \gamma ^{ab}\phi _\mu 
-2i\bar\varepsilon \gamma ^{[a}\hatR_\mu {}^{b]}(Q)-i\bar\varepsilon \gamma _\mu \hatR^{ab}(Q)
+4i\bar\varepsilon \phi ^{[a}(Q)e_\mu {}^{b]}\nn
&&-2i\bar\varepsilon \gamma ^{abcd}\psi _\mu v_{cd}-2i\bar\eta \gamma ^{ab}\psi _\mu 
-4\xi _K{}^{[a}e_\mu {}^{b]},\nn
\delta V_\mu ^{ij}&=&-6i\bar\varepsilon ^{(i}\phi ^{j)}_\mu 
-2i\bar\varepsilon ^{(i}\gamma ^a\hatR_{a\mu }{}^{j)}(Q)
-\myfrac{i}4\bar\varepsilon ^{(i}\gamma _\mu \gamma \dt \hatR^{j)}(Q)\nn
&&+4i\bar\varepsilon ^{(i}\gamma \dt v\psi ^{j)}_\mu 
-\myfrac{i}4\bar\varepsilon ^{(i}\gamma _\mu \chi ^{j)}+6i\bar\eta ^{(i}\psi ^{j)}_\mu ,\nn
\delta v_{ab}&=&\myfrac{i}8\bar\varepsilon \gamma _{ab}\chi 
-\myfrac{i}8\bar\varepsilon \gamma ^{cd}\gamma _{ab}\hatR_{cd}(Q)
+\myfrac{i}2\bar\varepsilon \hatR_{ab}(Q),\nn
\delta \chi ^i&=&D\varepsilon ^i-2\gamma ^c\gamma ^{ab}\varepsilon ^i\hatD_av_{bc}+\gamma \dt \hatR(U)^i{}_j\varepsilon ^j\nn
&&-2\gamma ^a\varepsilon ^i\epsilon _{abcde}v^{bc}v^{de}+4\gamma \dt v\eta ^i,\nn
\delta D&=&-i\bar\varepsilon \slashD\chi -\myfrac{i}2\bar\varepsilon \gamma \dt v\gamma \dt\hatR(Q)
-8i\bar\varepsilon \hatR_{ab}(Q)v^{ab}+i\bar\eta \chi ,
\end{eqnarray}
where the derivative $\calD_\mu $ is covariant only with respect to the
homogeneous transformations $\XM_{ab},\XD$ and $\XU^{ij}$ (and the $\XG$
transformation for non-singlet fields under the Yang-Mills group $G$). We have also
written the transformation law of the spin connection for
convenience.
The algebra of the $\XQ$ and $\XS$ transformations takes the form
\begin{eqnarray}
{}[\delta _Q(\varepsilon _1),\,\delta _Q(\varepsilon _2)]&=&\delta _P(2i\bar\varepsilon _1\gamma _a\varepsilon _2)
+\delta _M(2i\bar\varepsilon _1\gamma ^{abcd}\varepsilon _2v_{ab})
+\delta _U(-4i\bar\varepsilon _1^i\gamma \dt v\varepsilon _2^j)\nn
&&+\delta _S\left(\begin{array}{c}
2i\bar\varepsilon _1\gamma ^a\varepsilon _2\phi _{ai}(Q)
+i\bar\varepsilon _{1(i}\gamma _{ab}\varepsilon _{2j)}K^{abj}(Q)\\
+\myfrac3{32}i\bar\varepsilon _1\varepsilon _2\chi _i
+\myfrac3{32}i\bar\varepsilon _1\gamma ^a\varepsilon _2\gamma _a\chi _i
-\myfrac1{32}i\bar\varepsilon _{1(i}\gamma ^{ab}\varepsilon _{2j)}\gamma _{ab}\chi ^j
\end{array}\right)\nn
&&+\delta _K\left(\begin{array}{c}
2i\bar\varepsilon _1\gamma ^b\varepsilon _2f_b{}^a(M)
+\myfrac1{12}i\bar\varepsilon _1^{(i}\gamma ^{abc}\varepsilon _2^{j)}\hatR_{bcij}(U)\\
+\myfrac{i}6\bar\varepsilon _1\gamma ^{abcd}\varepsilon _2\hatD_bv_{cd}
+\myfrac{i}2\bar\varepsilon _1\varepsilon _2\hatD_bv^{ab}\\
+\myfrac56i\bar\varepsilon _1\gamma ^a\varepsilon _2v\dt v
+\myfrac83i\bar\varepsilon _1\gamma ^b\varepsilon _2v_{bc}v^{ca}\\
-\myfrac16i\bar\varepsilon _1\gamma ^{abcde}\varepsilon _2v_{bc}v_{de}
\end{array}\right),\label{eq:QQcommutator}\\
{}[\delta _S(\eta ),\delta _Q(\varepsilon )]&=&\delta _D(-2i\bar\varepsilon \eta )
+\delta _M(2i\bar\varepsilon \gamma ^{ab}\eta )+\delta _U(-6i\bar\varepsilon ^{(i}\eta ^{j)})\nn
&&+\delta _K\left(-\myfrac56i\bar\varepsilon \gamma _{abc}\eta v^{ab}+i\bar\varepsilon \gamma ^b\eta 
v_{ab}\right),\label{eq:SQcommutator}
\end{eqnarray}
where the translation $\delta _P(\xi ^a)$ is understood to be essentially 
the general coordinate 
transformation $\delta _{\rm GC}(\xi ^\lambda )$:
\begin{eqnarray}
\delta _P(\xi ^a)&=&\delta _{\rm GC}(\xi ^\lambda )-\delta _A(\xi ^\lambda h_\lambda ^A).
\end{eqnarray}
On a covariant quantity $\Phi $ with only flat indices, 
$\delta _P(\xi ^a)$ acts as  the full covariant derivative:
\begin{equation}
\delta _P(\xi ^a)\Phi =\xi ^a\left(\partial _a-\delta _A(h_a^A)\right)\Phi \equiv \xi ^a\hatD_a\Phi .
\end{equation}
Note the consistency that the quantities $\phi _a^i(Q)$ and
$f_a{}^b(M)$ on the
right-hand side of the algebra (\ref{eq:QQcommutator}) cancel out the
$S$ and $K$ gauge fields contained in $\delta _P(\xi ^a)$.

\section{ Transformation laws of matter multiplets}
\begin{table}[tb]
\caption{Matter multiplets in 5D.}
\label{table:5DM}
\begin{center}
\begin{tabular}{ccccc}\hline \hline
field      & type      &  remarks & {\it SU}$(2)$&  Weyl-weight    \\ \hline 
\multicolumn{5}{c}{Vector multiplet} \\ \hline
$W_\mu$      &  boson    & real gauge field   &  \bf{1}    &   0     \\
$M$& boson & real& \bf{1} & 1 \\ 
$\Omega ^i$& fermion  &{\it SU}$(2)$-Majorana  & \bf{2} &$\frac{3}{2}$ \\  
$Y_{ij}$    &  boson    & $Y^{ij}=Y^{ji}=(Y_{ij})^*$   & \bf{3} & 2 \\ \hline
\multicolumn{5}{c}{Hypermultiplet} \\ \hline
\Tspan{$\calA_i^\alpha$}     &  boson & 
$\calA^i_\alpha=\epsilon^{ij}\calA_j^\beta\rho_{\beta\alpha}=-(\calA_i^\alpha)^*$ &\bf{2}& $\frac{3}{2}$\\
$\zeta^\alpha$    &  fermion  & $\bar\zeta^\alpha\equiv(\zeta_\alpha)^\dagger\gamma_0 = \zeta^{\alpha\T}C$ 
& \bf{1}  & 2 \\ 
${\cal F}_i^\alpha$  &  boson    & $\calF_i^\alpha \equiv \alpha \XZ\calA_i^\alpha $,\,
${\cal F}^i_\alpha=-({\cal F}_i^\alpha)^*$  &  \bf{2}
&$\frac{5}{2}$\\
\hline
\multicolumn{5}{c}{Linear multiplet} \\ \hline
\Tspan{$L^{ij}$}& boson & $L^{ij}=L^{ji}=(L_{ij})^*$  &  \bf{3}   & 3 \\ 
$\varphi^i$ &  fermion  & {\it SU}$(2)$-Majorana & \bf{2}&$\frac{7}{2}$ \\  
$E_a$ & boson & real,\quad constrained by (\ref{eq:Con.E})& \bf{1} & 4 \\ 
$N$ & boson & real & \bf{1} & 4 \\ \hline
\multicolumn{5}{c}{Nonlinear multiplet} \\ \hline
\Tspan{$\Phi^i_\alpha$}  &  boson    & $SU(2)$ -valued
  &  \bf{2} & 0 \\ 
$\lambda^i$   &  fermion  & {\it SU}$(2)$-Majorana & \bf{2} &$\frac{1}{2}$ \\  
$V^a$    &  boson    & real & \bf{1} & 1 \\ 
$V^5$   & boson & real & \bf{1} & 1 \\ \hline
\end{tabular}
\end{center}
\end{table}
In 5D there are four kinds of multiplets: a vector multiplet, 
hypermultiplet, linear multiplet and nonlinear multiplet.
The components of the matter multiplets and their properties are listed
in Table \ref{table:5DM}. 
The tensor multiplet in 6D reduces to a vector multiplet in 5D with
constraints, and solving these constraints gives rise to an alternative 
type of the Weyl multiplet containing the two-form gauge field 
$B_{\mu \nu }$\cite{Nis-Raj} in the same way as in 6D. 
We discuss $B_{\mu \nu }$ in \S \ref{sec:another}.
  
The supersymmetry transformation laws of the matter multiplets are almost
identical to those obtained in Ref.~\citen{Kug-Oha1} 
in the Poincar\'e supergravity case 
if the `central charge vector multiplet' components are omitted in the 
latter.

\subsection{Vector multiplet}
An important difference between the vector multiplets in 5D and in 6D 
is the existence of the scalar component $M$ in 5D, which allows for the
introduction of the `very special geometry'\cite{Gun-Zag} $c_{IJK}M^IM^JM^K=1$ 
in the Poincar\'e supergravity theory.  
All the component fields of this multiplet are Lie-algebra valued,
e.g., $M$ is a matrix $M^\alpha {}_\beta =M^I(t_I)^\alpha {}_\beta $, where the $t_I$ are 
(anti-hermitian) generators of the gauge group $G$.
The $\XQ$ and $\XS$ transformation laws of the vector multiplet are given by
\begin{eqnarray}
\delta W_\mu &=&-2i\bar\varepsilon \gamma _\mu \Omega +2i\bar\varepsilon \psi _\mu M,\nn
\delta M&=&2i\bar\varepsilon \Omega ,\nn
\delta \Omega ^i&=&-\myfrac14\gamma \dt \hat F(W)\varepsilon ^i
-\myfrac12\slashD M \varepsilon ^i+Y^i{}_j\varepsilon ^j-M\eta ^i,\nn
\delta Y^{ij}&=&2i\bar\varepsilon ^{(i}\slashD\Omega ^{j)}-i\bar\varepsilon ^{(i}\gamma \dt v\Omega ^{j)}
-\myfrac{i}4\bar\varepsilon ^{(i}\chi ^{j)}M
-\myfrac{i}4\bar\varepsilon ^{(i}\gamma \dt \hatR^{j)}(Q)M\nn
&&-2ig\bar\varepsilon ^{(i}[M,\Omega ^{j)}]-2i\bar\eta ^{(i}\Omega ^{j)}.
\end{eqnarray}
The gauge group $G$ can be regarded as a subgroup of the
superconformal group, and the above transformation law of the gauge
field $W_\mu $ provides us with the additional structure functions, 
$f_{PQ}{}^G$ and $f_{QQ}{}^G$. For instance, the commutator of the two
$\XQ$ transformations becomes
\begin{equation}
[\delta _Q(\varepsilon _1),\,\delta _Q(\varepsilon _2)]=({\rm R.H.S.~of~(\ref{eq:QQcommutator})})
+\delta _G(-2i\bar\varepsilon _1\varepsilon _2M).\label{eq:QQcommutator2}
\end{equation}

For the reader's convenience, we give here
the transformation laws of the covariant
derivative of the scalar $\hatD_aM$ and the field strength $\hat F_{ab}(W)$:
\begin{eqnarray}
\delta \hatD_aM&=&2i\bar\varepsilon \hatD_a\Omega -2i\bar\varepsilon \phi _a(Q)M+i\bar\varepsilon \gamma _{abc}\Omega v^{bc}
+2ig\bar\varepsilon \gamma _a[\Omega ,\,M]+2i\bar\eta \gamma _a\Omega +2\xi _{Ka}M,\nn
\delta \hat F_{ab}(W)&=&4i\bar\varepsilon \gamma _{[a}\hatD_{b]}\Omega 
-2i\bar\varepsilon \gamma _{cd[a}\gamma _{b]}\Omega v^{cd}+2i\bar\varepsilon \hatR_{ab}(Q)M
-4i\bar\eta \gamma _{ab}\Omega .
\end{eqnarray}
The transformation laws of a matter field acted on by a covariant
derivative and
the supercovariant curvature (field strength) are derived easily using the
simple fact that the transformation of any covariant quantity also
gives a covariant quantity and hence cannot contain gauge fields
explicitly; that is, gauge fields can appear only implicitly in the
covariant derivative or in the form of supercovariant curvatures,
as long as the algebra closes. Similarly, the Bianchi identities can be
computed by discarding the naked gauge fields with no derivative,
because both sides of the 
identity are, of course, covariant. For example, we have 
\begin{equation}
\hatD_{[a}\hat F_{bc]}(W)=-2i\bar\Omega \gamma _{[a}\hatR_{bc]}(Q).
\end{equation}

\subsection{Hypermultipet}
The hypermultiplet in 5D consists of scalars $\calA^i_\alpha $,
spinors $\zeta _\alpha $ and auxiliary fields $\calF^i_\alpha $. They carry the
index $\alpha ~(=1,2,\ldots,2r)$ of the representation of a subgroup $G'$ of the gauge group $G$, 
which is 
raised (or lowered) with a $G'$ invariant tensor 
 $\rho _{\alpha \beta }$ (and $\rho ^{\alpha \beta }$ with $\rho ^{\gamma \alpha }\rho _{\gamma \beta }=\delta _\beta ^\alpha $) 
like $\calA^i_\alpha =\calA^{i\beta }\rho _{\beta \alpha }$.
This multiplet gives an
infinite dimensional representation of a central charge gauge group
$U_Z(1)$, which we regard as a subgroup of 
the group $G$. 
The scalar fields $\calA^i_\alpha $ satisfy the reality condition
\begin{eqnarray}
\calA_\alpha ^i&=&\epsilon ^{ij}\calA^\beta _j\rho _{\beta \alpha }=-(\calA_i^\alpha )^*,
\quad \calA_{i\alpha }=(\calA^{i\alpha })^*,
\end{eqnarray}
and the tensor $\rho _{\alpha \beta }$ can generally be brought into the standard
form $\rho =\epsilon \otimes {\bf 1}_r$ by a suitable field
redefinition. Therefor
$\calA^i_\alpha $ can be identified with $r$ quaternions.
 Thus the group $G'$ acting linearly on the hypermultiplet should be a subgroup
of $GL(r;\XH )$:
\begin{eqnarray}
&&\delta_{G'}(t)\calA^\alpha_i=gt^\alpha{}_\beta\calA^\beta_i, \qquad  
\delta_{G'}(t)\calA^i_\alpha=g(t^\alpha{}_\beta)^*\calA^i_\beta=-gt_\alpha{}^\beta\calA_\beta^i, \nn
&&t_\alpha{}^\beta\equiv\rho_{\alpha\gamma}t^\gamma{}_\delta\rho^{\delta\beta}=-(t^\alpha{}_\beta)^*.
\label{eq:H.Gtr}
\end{eqnarray}
Note that the spinors $\zeta _\alpha $ do not satisfy
the $SU(2)$-Majorana condition
explicitly, but rather the $USp(2r)$-Majorana condition,
\begin{eqnarray}
\bar\zeta ^\alpha &\equiv &(\zeta _\alpha )^\dagger \gamma ^0=\rho ^{\alpha \beta }(\zeta _\beta )^\T C=(\zeta ^\alpha )^{\T}C.
\end{eqnarray}

The $\XQ$ and $\XS$ transformations of the $\calA^i_\alpha $ and $\zeta _\alpha $
are given by
\begin{eqnarray}
\delta \calA^i_\alpha &=&2i\bar\varepsilon ^i\zeta _\alpha ,\nn
\delta \zeta ^\alpha &=&\slashD\calA^\alpha _j\varepsilon ^j-\gamma \dt v\varepsilon ^j\calA^\alpha _j
-M_*\calA^\alpha _j\varepsilon ^j+3\calA^\alpha _j\eta ^j,
\end{eqnarray}
and with these rules, to realize the superconformal algebra on the
hypermultiplet requires the
following two $\XQ$  and $\XS$ invariant constraints:
\begin{eqnarray}
0&=&\slashD\zeta ^\alpha +\myfrac12\gamma \dt v\zeta ^\alpha 
-\myfrac18\chi ^i\calA^\alpha _i+\myfrac38\gamma \dt \hatR^i(Q)\calA^\alpha _i\nn
&&+M_*\zeta ^\alpha -2\Omega ^i_*\calA^\alpha _i,\nn
0&=&-\hatD^a\hatD_a\calA^\alpha _i+M_*M_*\calA^\alpha _i
+4i\bar\Omega _{i*}\zeta ^\alpha -2Y_{ij*}\calA^{\alpha j}\nn
&&-\myfrac{i}4\bar\zeta ^\alpha \left(\chi +\gamma \dt \hatR(Q)\right)
+\left(\myfrac{3}{16}\hatR(M)+\myfrac18D-\myfrac14v^2\right)\calA^\alpha _i,\label{eq:con.H}
\end{eqnarray}
where $\theta _*= M_*,\Omega _*,\ldots$ represents the Yang-Mills
transformations with parameters
$\theta $, including the central charge transformation,
that is $\delta _G(\theta )=\delta _{G'}(\theta )+\delta _Z(\theta^0)$.
$\XZ$ denotes the
generator of the $U_Z(1)$ transformation and $\XV^0=(V_\mu ^0\equiv A_\mu ,M^0\equiv \alpha ,\Omega 
^0,Y^{0ij})$ denotes the $U_Z(1)$ vector multiplet. For example,
acting on the scalar $\calA^\alpha _i$, we have
\begin{equation}
\quad M_*\calA^\alpha _i=gM^\alpha {}_\beta \calA^\beta _i+\alpha \XZ\calA^\alpha _i.
\end{equation}
The hypermultiplet in 6D exists only as an on-shell multiplet,
since constraints similar to (\ref{eq:con.H}) are equations of
motion there. Here in 5D, however, it becomes an off-shell multiplet,
as explained in Ref.~\citen{Kug-Oha1}.

First, there appears no constraint on the first $U_Z(1)$
transformation of $\calA^\alpha _i$, so it defines the auxiliary field
\begin{equation}
\calF^\alpha _i\equiv \alpha \XZ\calA^\alpha _i,
\end{equation}
which is necessary for closing the algebra off-shell and balancing the 
numbers of boson and fermion degrees of freedom.
Next, there are the undefined $U_Z(1)$ transformations $\XZ\zeta _\alpha ,
\XZ(\XZ\calA^\alpha _i)~(=\alpha ^{-1}\XZ\calF^\alpha _i)$ in the constraints
(\ref{eq:con.H}), and therefor we do not interpret the constraints as the
equation of motion but as definitions of these $U_Z(1)$
transformations.
The first constraint of (\ref{eq:con.H}), for example, 
gives the $U_Z(1)$ transformation of the spinor $\zeta _\alpha $ as 
\begin{eqnarray}
{\XZ}\zeta ^\alpha &=&-{\alpha +\gamma ^aA_a\over \alpha ^2-A^aA_a}
\Bigr(\slashD'\zeta ^\alpha +\myfrac12\gamma \dt v\zeta ^\alpha 
-\myfrac18\chi ^i\calA^\alpha _i+\myfrac38\gamma \dt \hatR^i(Q)
\calA^\alpha _i.\nn
&&\qquad \qquad \qquad \qquad 
+gM^\alpha {}_\beta\zeta ^\beta -2g\Omega ^{i\alpha}{}_\beta\calA^\beta _i-\myfrac2\alpha \Omega ^{0i}\calF^\alpha _i\Bigr).
\end{eqnarray}
Note that $\hatD_a\zeta _\alpha $ contains the
$U_Z(1)$ covariantization $-\delta _Z(A_a)
\zeta _\alpha $ and $\hatD_a'$ denotes a covariant derivative with the $-\delta _
Z(A_a)$ term omitted. Also, the second constraint gives the
$U_Z(1)$ transformation of $\calF^\alpha _i$, which we do not show
explicitly here. Finally, the $\XQ$ and $\XS$ transformations of the
auxiliary field $\calF^\alpha _i$ are given by requiring that the $U_Z(1)$
transformation commute with the $\XQ$ and $\XS$
transformations on $\calA^\alpha _i$: 
\begin{eqnarray}
\delta \calF^i_\alpha &=&\delta \left(\delta _Z(\alpha )\calA^\alpha _i\right)
=\left(\delta _Z(\alpha )\delta +\delta _Z(\delta \alpha )\right)\calA^\alpha _i\nn
&=&2i\bar\varepsilon ^i(\alpha \XZ\zeta _\alpha )+\myfrac{2i}\alpha \bar\varepsilon \Omega ^0\calF^i_\alpha .
\end{eqnarray}

\subsection{Linear multiplet\label{sec:linear}}
The linear multiplet consists of the components listed in 
Table \ref{table:5DM} and may generally carry a non-Abelian charge of
the gauge group $G$. This multiplet, apparently, contains 9 Bose 
and 8 Fermi fields, so that the closure of the algebra on this multiplet 
requires the constraint (\ref{eq:Con.E}), which can be solved in terms of 
a three-form gauge field $E_{\mu \nu \lambda }$. A four-form gauge 
field $H_{\mu \nu \rho \sigma }$ can also be introduced for rewriting the scalar
component of this multiplet.

The $\XQ$ and $\XS$ transformation laws of the linear multiplet are
given by
\begin{eqnarray}
\delta L^{ij}&=&2i\bar\varepsilon ^{(i}\varphi ^{j)},\nn
\delta \varphi ^i&=&-\slashD L^{ij}\varepsilon _j+\myfrac12\gamma ^a\varepsilon ^iE_a
+\myfrac12\varepsilon ^iN\nn
&&+2\gamma \dt v\varepsilon _jL^{ij}+gML^{ij}\varepsilon _j-6L^{ij}\eta _j,\nn
\delta E^a&=&2i\bar\varepsilon \gamma ^{ab}\hatD_b\varphi 
-2i\bar\varepsilon \gamma ^{abc}\varphi v_{bc}
+6i\bar\varepsilon \gamma _b\varphi v^{ab}
+2i\bar\varepsilon ^i\gamma ^{abc}\hatR^j_{bc}(Q)L_{ij}\nn
&&+2ig\bar\varepsilon \gamma ^aM\varphi -4ig\bar\varepsilon ^i\gamma ^a\Omega ^jL_{ij}
-8i\bar\eta \gamma _a\varphi, \nn
\delta N&=&-2i\bar\varepsilon \slashD\varphi -3i\bar\varepsilon \gamma \dt v\varphi 
+\myfrac12i\bar\varepsilon ^i\chi ^jL_{ij}
-\myfrac32i\bar\varepsilon ^{(i}\gamma \dt \hatR^{j)}(Q)L_{ij}\nn
&&+4ig\bar\varepsilon ^{(i}\Omega ^{j)}L_{ij}-6i\bar\eta \varphi. \label{eq:trf.L}
\end{eqnarray}
The algebra closes if $E^a$ satisfies the following $Q$ and
$S$ invariant constraint:
\begin{equation}
\hatD_aE^a+i\bar\varphi \gamma \dt \hatR(Q)+gMN+4ig\bar\Omega \varphi
+2gY^{ij}L_{ij}=0.
\label{eq:Con.E}
\end{equation}
This constraint can be separated into two parts, a total
derivative part and the part proportional to 
the Yang-Mills coupling $g$: 
\begin{equation}
e^{-1}\partial _\lambda (e{\cal V}^\lambda )+2ge^{-1}{\cal H}_{VL}=0,\label{eq:Con.V}
\end{equation}
where, ${\cal V}^a$ and ${\cal H}_{VL}$ are given by
\begin{eqnarray}
{\cal V}^a&=&E^a-2i\bar\psi _b\gamma ^{ba}\varphi +2i\bar\psi _b\gamma ^{abc}L\psi _c,\nn
e^{-1}{\cal H}_{VL}&=&Y^{ij}L_{ij}+2i\bar\Omega \varphi
 +2i\bar\psi ^a_i\gamma _a\Omega _jL^{ij}-\myfrac12W_a{\cal V}^a\nn
&&+\myfrac12M\left(N-2i\bar\psi _b\gamma ^{b}\varphi 
-2i\bar\psi _a^{(i}\gamma ^{ab}\psi _b^{j)}L_{ij}\right).\label{eq:def.VH}
\end{eqnarray}
When the linear multiplet is inert under the
$\XG$ transformation, that is $g=0$, 
this constraint can be solved in terms of 
a three-form gauge field $E_{\mu \nu \lambda }$ as ${\cal V}^\lambda =e^{-1}\epsilon ^{\lambda \mu 
\nu \rho \sigma }\partial _\mu E_{\nu \rho \sigma }/6$,
which possesses the additional gauge symmetry 
$\delta _E(\Lambda )E_{\mu \nu \lambda }=
3\partial _{[\mu }\Lambda_{\nu \lambda ]}$.
Hence the linear multiplet becomes
an unconstrained multiplet $(E_{\mu \nu \lambda },L^{ij},\varphi ^i,N)$. 

It shoud be noted that
in 6D, the linear multiplet
requires a similar constraint on the 6D vector $E^a$, 
and this constraint can be solved in
terms of the four-form gauge field $E_{\mu \nu \rho \sigma }$
in a similar manner.
This 6D four-form field yields a three-form field $E_{\mu \nu \lambda }$ and 
a four-form field $H_{\mu \nu \rho \sigma }$
through the simple reduction, 
while the 6D vector $E^a$ reduces to the 5D vector $E^a$
and the scalar $N$.
Thus we expect that the scalar field $N$ can be rewritten in terms of 
a four-form field $H_{\mu \nu \rho \sigma }$ in this 5D linear multiplet.
(Note the number of degrees of freedom of the $H_{\mu \nu \rho \sigma}$
is 1 in
5D.) The quantity ${\cal H}_{VL}$
contains $N$, and any transformation of this quantity becomes a total
derivative, because the constraint (\ref{eq:Con.V}) is invariant under
the full transformations. Thus it
can be rewritten with $H_{\mu \nu \rho \sigma }$ in 
the form
\begin{eqnarray}
2{\cal H}_{VL}&=&-\myfrac1{4!}\epsilon ^{\lambda \mu \nu \rho \sigma }\partial _\lambda (H_{\mu \nu \rho \sigma }
-4W_\mu E_{\nu \rho \sigma }), \label{eq:Nreplace}
\end{eqnarray}
where the extra term $W_{[\mu }E_{\nu \rho \sigma ]}$ on
the right hand-side is
inserted for later convenience.  
With this rewriting, the constraint (\ref{eq:Con.V}) can be solved even
for the case that the linear
multiplet carries a charge of the gauge group $G$. Indeed, since 
the r.h.s. of (\ref{eq:Nreplace}) is a total derivative, we have
\begin{eqnarray}
&&e^{-1}\partial _\lambda \left\{e{\cal V}^\lambda 
-\myfrac{g}{4!}\epsilon ^{\lambda \mu \nu \rho \sigma }(H_{\mu \nu \rho \sigma }
-4W_\mu E_{\nu \rho \sigma })\right\}=0\nn
&&\rightarrow \quad {\cal V}^\lambda =\myfrac1{4!}e^{-1}\epsilon ^{\lambda \mu \nu \rho \sigma }
\left(4\partial _{[\mu }E_{\nu \rho \sigma ]}-4gW_\mu E_{\nu \rho \sigma }+gH_{\mu \nu \rho \sigma }\right).
\label{eq:Ereplace}
\end{eqnarray}
The transformation laws of $E_{\mu \nu \lambda }$ and $H_{\mu \nu \rho \sigma }$ must be determined
up to the additional gauge symmetry
$\delta _H(\Lambda )H_{\mu \nu \rho \sigma }=
4(\partial _{[\mu }-gW_{[\mu })\Lambda _{\nu \rho \sigma ]},\,\delta _H(\Lambda )E_{\mu \nu \lambda }=-g\Lambda _{\mu \nu \lambda }$,
so that the all transformation laws of both sides of
(\ref{eq:Ereplace})
are the same for consistency. Also the transformation laws of the
tensor gauge fields defined in this way are consistent with the
replacement equation (\ref{eq:Nreplace}), because this equation is
satisfied automatically due to the invariant equations
(\ref{eq:Con.V}) and (\ref{eq:Ereplace}). Now, let us rewrite 
the replacement equation of $N$, (\ref{eq:Nreplace}), and the solution of 
$E^a$, (\ref{eq:Ereplace}), into the following two invariant equations:  
\begin{eqnarray}
E^a&=&\myfrac1{4!}\epsilon ^{abcde}\hat F_{bcde}(E),
\nn
MN+2Y_{ij}L^{ij}+4i\bar\Omega \varphi&=&
-\myfrac1{5!}\epsilon ^{abcde}\hat F_{abcde}(H).\label{eq:CovEq}
\end{eqnarray}
The quantities $\hat F_{abcd}(E)$ and $\hat F_{abcde}(H)$
are the field strengths given by
\begin{eqnarray}
\hat F_{\mu \nu \rho \sigma }(E)&=&4D_{[\mu }E_{\nu \rho \sigma ]}+gH_{\mu \nu \rho \sigma }+
8i\bar\psi _{[\mu }\gamma _{\nu \rho \sigma ]}\varphi 
+24i\bar\psi _{[\mu }^i\gamma _{\nu \rho }\psi _{\sigma ]}^jL_{ij},\nn
\hat F_{\lambda \mu \nu \rho \sigma }(H)&=&5D_{[\lambda }H_{\mu \nu \rho \sigma ]}-10F_{[\lambda \mu }(W)E_{\nu \rho \sigma ]}\nn
&&-10i\bar\psi _{[\lambda }\gamma _{\mu \nu \rho \sigma ]}M\varphi 
+20i\bar\psi ^i_{[\lambda }\gamma _{\mu \nu \rho \sigma ]}\lambda ^jL_{ij}
-40i\bar\psi ^i_{[\lambda }\gamma _{\mu \nu \rho }\psi ^j_{\sigma ]}ML_{ij},\nn
\end{eqnarray}
where the derivative $D_\mu $ is covariant with respect to 
the $G$ transformation:$D_\mu \equiv \partial _\mu -gW_\mu $.
The transformation laws of
$E_{\mu \nu \lambda }$ and $H_{\mu \nu \rho \sigma }$ can be
understand from 
the fact that the left-hand sides of the equations in (\ref{eq:CovEq}) 
are covariant under the full transformation, and so the field
strengths on the r.h.s. must also be fully covariant. 
With $\delta \equiv \delta _Q(\varepsilon )+\delta _S(\eta )+\delta _G(\Lambda )+\delta _E(\Lambda _{\mu \nu })+\delta _H(\Lambda _{\mu \nu \lambda })$,
we have
\begin{eqnarray}
\delta E_{\mu \nu \lambda }&=&3D_{[\mu }\Lambda _{\nu \lambda ]}+g\Lambda E_{\mu \nu \lambda }-g\Lambda _{\mu \nu \lambda }
-2i\bar\varepsilon \gamma _{\mu \nu \lambda }\varphi 
-12i\bar\varepsilon ^i\gamma _{[\mu \nu }\psi ^j_{\lambda ]}L_{ij},\nn
\delta H_{\mu \nu \rho \sigma }&=&4D_{[\mu }\Lambda _{\nu \rho  \sigma ]}
+g\Lambda H_{\mu \nu \rho \sigma }+6F_{[\mu \nu }(W)\Lambda _{\rho \sigma ]}\nn
&&+2i\bar\varepsilon \gamma _{\mu \nu \rho \sigma }M\varphi 
-4i\bar\varepsilon ^i\gamma _{\mu \nu \rho \sigma }\Omega ^jL_{ij}\nn
&&+16i\bar\varepsilon ^i\gamma _{[\mu \nu \rho }\psi ^j_{\sigma ]}ML_{ij}+4(\delta _Q(\varepsilon )W_{[\mu })E_{\nu \rho \sigma ]}.
\end{eqnarray}
These transformation laws are truly consistent with
(\ref{eq:CovEq}), and thus with (\ref{eq:Ereplace}). With these laws,
the following modified algebra closes on the tensor gauge fields $E_
{\mu \nu \lambda }$ and $H_{\mu \nu \rho \sigma }$:
\begin{eqnarray}
[\delta _Q(\varepsilon _1),\,\delta _Q(\varepsilon _2)]&=&
({\rm R.H.S.~of~(\ref{eq:QQcommutator2})})
+\delta _E(4i\bar\varepsilon _1^i\gamma _{\mu \nu }\varepsilon _2^jL_{ij})\nn
&&+\delta _H\left(4i\bar\varepsilon _1^i\gamma _{\mu \nu \lambda }\varepsilon _2^jML_{ij}
-2i\bar\varepsilon _1\varepsilon _2ME_{\mu \nu \lambda }\right),\nn
{}[\delta _Q(\varepsilon ),\,\delta _E(\Lambda _{\mu \nu })]&=&\delta _H\left(3\delta _Q(\varepsilon )W_{[\mu }\Lambda _{\nu \lambda ]}
\right),\nn
{}[\delta _G(\Lambda ),\,\delta _E(\Lambda _{\mu \nu })]&=&\delta _E(-g\Lambda \Lambda _{\mu \nu }),\qquad 
{}[\delta _G(\Lambda ),\,\delta _H(\Lambda _{\mu \nu \lambda })]=\delta _H(-g\Lambda \Lambda _{\mu \nu \lambda }).\nn
\end{eqnarray}
  This fact also justifies the replacement (\ref{eq:Nreplace}) 
  algebraically. Nevertheless, in order to actually claim that 
$(E_{\mu \nu \lambda },\,H_{\mu \nu \rho \sigma },\,L^{ij},\ \varphi ^i)$ 
gives a new version of the linear multiplet, 
we must show that the component $N$ can be expressed in terms of 
$H_{\mu \nu \rho \sigma }$ by solving (\ref{eq:Nreplace}).
The point is that the left hand side of (\ref{eq:Nreplace}) ${\cal
  H}_{VL}$ contains $N$ in the form $M N$, but
  the Lie-algebra valued scalar $M$ is, of course,
  not always invertible.
 In some particular cases, the matrix $M$ can be invertible. For
  example, the determinant of the $SU(2)$-valued matrix $M^a(\sigma _a/2)$
  does not vanish in the domain $\sum_{a=1}^3M^aM^a\not=0$.
Therefor, the linear multiplet can take the doublet representation of 
  $SU(2)$ as a subgroup  of $G$ in this domain.
 
\subsection{Nonlinear multiplet} 
A nonlinear multiplet is a multiplet whose component fields are
transformed nonlinearly. The first component, $\Phi ^i_\alpha $, carries an
additional gauge-group $SU(2)$ index $\alpha ~(=1,2)$, as well as the
superconformal $SU(2)$ index $i$. The index $\alpha $ is also raised (and
lowered) by using the invariant tensor $\epsilon ^{\alpha \beta }$ 
(and $\epsilon _{\alpha \beta }$ with $\epsilon ^{\gamma \alpha }\epsilon _{\gamma \beta }=
\delta ^\alpha _\beta $)   as $\Phi ^i_\alpha =\Phi ^{i\beta }\epsilon _{\beta \alpha }$.
The field $\Phi ^i_\alpha $ takes values in $SU(2)$ and hence satisfies
\begin{eqnarray}
\Phi ^i_\alpha \Phi ^\alpha _j=\delta ^i_j,\qquad \Phi ^\alpha _i\Phi ^i_\beta =\delta ^\alpha _\beta .
\end{eqnarray}
  
The $Q$, $S$ and $K$ transformation laws of this multiplet are given by
\begin{eqnarray}
\delta \Phi _i^\alpha &=&2i\bar\varepsilon _{(i}\lambda _{j)}\Phi ^{j\alpha },\nn
\delta \lambda ^i
&=&-\Phi _\alpha ^i\slashD\Phi ^\alpha _j\varepsilon ^j+M_{\alpha \beta }\Phi ^{\alpha i}\Phi ^{\beta j}\varepsilon _j
+\gamma \dt v\varepsilon ^i\nn
&&+\myfrac12\gamma ^aV_a\varepsilon ^i+\myfrac12V^5\varepsilon ^i
-2i\bar\varepsilon ^j\lambda ^i\lambda _j-3\eta ^i,\nn
\delta V_a&=&2i\bar\varepsilon \gamma _{ab}\hatD^b\lambda 
-i\bar\varepsilon \gamma _a\gamma \dt V\lambda +i\bar\varepsilon \gamma _a\lambda V^5\nn
&&+2i\bar\varepsilon \gamma ^b\lambda v_{ab}
+\myfrac14i\bar\varepsilon \gamma _a\chi 
+2i\bar\varepsilon \gamma ^b\hatR_{ab}(Q)-i\bar\varepsilon \gamma _a\gamma \dt \hatR(Q)\nn
&&+2i\bar\varepsilon ^i\gamma _a\Phi _{\alpha i}\slashD\Phi ^\alpha _j\lambda ^j
-4ig\bar\varepsilon ^i\gamma _a\Omega ^j_{\alpha \beta }\Phi ^\alpha _i\Phi ^\beta _j
-2ig\bar\varepsilon ^i\gamma _a\lambda ^jM_{\alpha \beta }\Phi ^\alpha _i\Phi ^\beta _j\nn
&&-2i\bar\eta \gamma _a\lambda -6\xi _{Ka},\nn
\delta V^5&=&-2i\bar\varepsilon \slashD\lambda +i\bar\varepsilon \gamma \dt V\lambda -i\bar\varepsilon \lambda V^5
-i\bar\varepsilon \gamma \dt v\lambda -\myfrac14i\bar\varepsilon \chi 
+\myfrac34i\bar\varepsilon \gamma \dt \hatR(Q)\nn
&&-2i\bar\varepsilon ^i\Phi _{\alpha i}\slashD\Phi ^\alpha _j\lambda ^j
+4ig\bar\varepsilon \Omega ^j_{\alpha \beta }\Phi ^\alpha _i\Phi ^\beta _j
+2ig\bar\varepsilon ^i\lambda ^jM_{\alpha \beta }\Phi ^\alpha _i\Phi ^\beta _j.\nn
\end{eqnarray}
As in the linear multiplet case, the nonlinear multiplet also needs
the following $Q$, $S$ and $K$ invariant constraint for the closure
of the algebra:
\begin{eqnarray}
&&\hatD^aV_a-\myfrac12V_aV^a+\myfrac12(V^5)^2
+\hatD^a\Phi ^i_\alpha \hatD_a\Phi ^\alpha _i+2i\bar\lambda \slashD\lambda \nn
&&+2i\bar\lambda ^i\Phi _{\alpha i}\slashD\Phi ^\alpha _j\lambda ^j+i\bar\lambda \gamma \dt v\lambda \nn
&&+\myfrac38\hatR(M)+\myfrac14D-\myfrac12v^2
+\myfrac{i}2\bar\lambda \chi -\myfrac{i}2\bar\lambda \gamma \dt\hatR(Q)\nn
&&+2gY^{ij}_{\alpha \beta }\Phi ^\alpha _i\Phi ^\beta _j-8ig\bar\lambda ^i\Omega ^j_{\alpha \beta }\Phi ^\alpha _i\Phi ^\beta _j
-2ig\bar\lambda ^i\lambda ^jM_{\alpha \beta }\Phi ^\alpha _i\Phi ^\beta _j\nn
&&+g^2M^\alpha _\beta M^\beta _\alpha =0.
\end{eqnarray}
This constraint can be solved for the scalar of the Weyl multiplet
$D$, and this solution presents us with a new (40+40) Weyl multiplet,
which possesses the unconstrained nonlinear multiplet instead of $D$. 


\section{Embedding and invariant action formulas}
\subsection{Embedding formulas}
We now give some embedding formulas that give a known type of
multiplet using a (set of) multiplet(s).

   To determine embedding formulas that give the linear multiplet $\XL$
by means of other multiplets is not difficult for the following reason.
When the transformation of the lowest component $L^{ij}$ of a
multiplet takes the form $2i\bar\varepsilon ^{(i}\varphi ^{j)}$, the
superconformal algebra consisting of
(\ref{eq:QQcommutator}) and (\ref{eq:SQcommutator})
demands that all the other higher components must uniquely transform
in the form given in Eq.~(\ref{eq:trf.L}) and that the constraint
(\ref{eq:Con.E}) should hold. Therefore, in order to identify all the
components of the linear multiplet, we have
only to examine the transformation law up 
to the second component $\varphi ^i$, as long as the algebra closes on
the embedded multiplets.      

The vector multiplets can be embedded into the linear multiplet with 
arbitrary quadratic homogeneous polynomials $f(M)$ 
of the first components $M^I$
of the vector multiplets.  The index $I$ labels the generators $t_I$ of the
gauge group $G$, which is generally non-simple. 
  These embedding formulas $\XL(\XV)$ are
\begin{eqnarray}
L_{ij}(\XV )&=& Y_{ij}^If_I-i\bar\Omega ^I_i\Omega ^J_jf_{IJ},\nn
\varphi _i(\XV )&=&-\myfrac14\left(\chi _i+\gamma \dt\hatR_i(Q)\right)f\nn
&&+\left(\slashD\Omega ^I_i-\myfrac12\gamma \dt v\Omega _i^I-g[M,\Omega ]^I\right)f_I\nn
&&+\left(-\myfrac14\gamma \dt\hat F^I(W)\Omega ^J+\myfrac12\slashD M^I\Omega ^J
-Y^I\Omega ^J\right)f_{IJ},\nn
E_a(\XV )&=&\hatD^b\left(4v_{ab}f+\hat F_{ab}^I(W)f_I
+i\bar\Omega ^I\gamma _{ab}\Omega ^Jf_{IJ}\right)\nn
&&+\left(-i\bar\Omega ^I\gamma _{abc}\hatR^{bc}(Q)-2ig[\bar\Omega ,\gamma _a\Omega ]^I+g[M,\hatD_aM]^I
\right)f_I\nn
&&+\left(-2ig\bar\Omega ^I\gamma _a[M,\Omega ]^J+\myfrac18\epsilon _{abcde}\hat
F^{bcI}(W)\hat F^{deJ}(W)\right)f_{IJ},\nn
N(\XV )&=&-\hatD^a\hatD_af
+\left(-\myfrac12 D +\myfrac14 \hat R(M)-3v\dt v\right)f\nn
&&+\left(-2\hat F_{ab}(W)v^{ab}+i\bar\chi \Omega ^I+2ig[\bar\Omega ,\Omega ]^I\right)f_I\nn
&&+\left(\begin{array}{c}
         -\myfrac14\hat F_{ab}^I(W)\hat F^{abJ}(W) 
         +\myfrac12\hatD^aM^I\hatD_aM^J\\
           +2i\bar\Omega ^I\slashD\Omega ^J-i\bar\Omega ^I\gamma \dt v\Omega ^J+Y^I_{ij}Y^{Jij}
\end{array}
\right)f_{IJ},\label{eq:VtoL}
\end{eqnarray}
where the commutator $[X,\,Y]^I$ represents 
$[X,Y]^It_I\equiv X^IY^J[t_I,\,t_J]$, and 
\begin{eqnarray}
f\equiv f(M),\quad f_I\equiv {\partial f\over \partial M^I},\quad f_{IJ}\equiv {\partial ^2f\over \partial M^I\partial M^J}.
\end{eqnarray}
Here, it is easy to see 
that we cannot generalize the function $f(M)$ further. 
For example,  
the lowest component $L_{ij}$ is $S$ invariant. Thus the right-hand side 
of the first equation of this formula has to be  
$S$ invariant, and this fact requires $M^If_I=2f$. Therefore $f(M)$
must be a homogeneous quadratic function of the scalar field $M$:
$f(M)=f_{IJ}M^IM^J/2$.

The product of the two
hypermultiplets $\XH=(\calA^i_\alpha ,\,\zeta _\alpha ,\,\calF^i_\alpha )$ 
and $\pr\XH=(\pr\calA^i_\alpha ,\,\pr\zeta _\alpha ,\,\pr\calF^i_\alpha )$ can also compose a linear
multiplet $\XL(\XH,\pr\XH)$ as follows:  
\begin{eqnarray}
L^{ij}_{\alpha \beta }(\XH,\pr\XH)&=&\calA^{(i}_\alpha \pr\calA^{j)}_\beta, \nn
\varphi ^i_{\alpha \beta }(\XH,\pr\XH)&=&\zeta _\alpha \pr\calA^i_\beta +\pr\zeta _\beta \calA^i_\alpha, \nn
E^a_{\alpha \beta }(\XH,\pr\XH)&=&\calA^i_\alpha \hatD^a\pr\calA_{\beta i}
+\pr\calA^i_\beta \hatD^a\calA_{\alpha i}
-2i\bar\zeta _\alpha \gamma ^a\pr\zeta _\beta, \nn
N_{\alpha \beta }(\XH,\pr\XH)&=&-\calA^i_\alpha M_*\pr\calA_{\beta i}
-\pr\calA^i_\beta M_*\calA_{\alpha i}
-2i\bar\zeta _\alpha \pr\zeta _\beta .\label{eq:HtoL}
\end{eqnarray}
Here, this linear multiplet transforms non-trivially under the $U_Z(1)$
transformation, in addition to the transformations that are self-evident 
from the index structure; e.g., $\delta _Z(\alpha )L^{ij}_{\alpha \beta }=\calF^{(i}_\alpha 
\pr\calA^{j)}_\beta +\calA^{(i}_\alpha \calF^{j)}_\beta $. For this multiplet,
therefore, the `group action terms', like $gML^{ij}$ appearing in the 
$Q$ transformation law (\ref{eq:trf.L}), and the action formula, which
we discuss in the next subsection, should be understood to contain
not only the usual gauge group action but also the $U(1)$ action:
 $gM\,\rightarrow \,M_*=\delta _{G'}(M)+\delta _Z(\alpha )$.  
Also note that $\XZ^n\XH$ with the arbitrary number $n$
can be substituted for $\XH$ and $\pr\XH$ in the above formulas,
because $\XZ\XH$ also transforms as a hypermultiplet.
 
 Conversely, we can also embed the linear multiplet into the vector
 multiplet. The following combination of the components of the
 linear multiplet is $S$ invariant and carries Weyl-weight $1$:
\begin{equation}
M(\XL)=  NL^{-1}+i\bar\varphi ^i\varphi ^jL_{ij}L^{-3},\label{eq:embedLtoV}
\end{equation}
with $L=\sqrt{L^{ij}L_{ij}}$.
This embedding formula is a non-polynomial function of the field,
and for this reason,
the embedding formulas for the higher components becomes quite complicated.    
Though we have not confirmed that the embedding (\ref{eq:embedLtoV}) is 
consistent with the transformation laws of the higher components, 
this formula agrees with the formula in the Poincar\'e
supergravity in 5D presented by Zucker \cite{Zuk3}
up to the components of 
the `central charge vector multiplet'. Thus it must be a correct form.

\subsection{Invariant action formula}
The quantity ${\cal H}_{VL}$ appearing in (\ref{eq:def.VH})
transforms into a total derivative under all of the gauge transformations
and has Weyl-weight $5$.
It therefor represents a possibility as the invariant
action formula.  However (\ref{eq:Con.V}) implies that 
${\cal H}_{VL}$ itself is a total derivative and so cannot give
 an action formula.
Fortunately, the invariant action formula can be found in the following way 
with a simple modification of the expression of ${\cal H}_{VL}$.
Let us consider the action formula
\begin{eqnarray}
e^{-1}{\cal L}(\XV\cdot\XL)
&\equiv &Y^{ij}\cdot L_{ij}+2i\bar\Omega \cdot \varphi
 +2i\bar\psi ^a_i\gamma _a\Omega _j\cdot L^{ij}\nn
&&-\myfrac12W_a\cdot \left(E^a-2i\bar\psi _b\gamma ^{ba}\varphi +2i\bar\psi _b^{(i}\gamma 
 ^{abc}\psi _c^{j)}L_{ij}\right)\nn
&&+\myfrac12M\cdot \left(N-2i\bar\psi _b\gamma ^{b}\varphi 
-2i\bar\psi _a^{(i}\gamma ^{ab}\psi _b^{j)}L_{ij}\right),\label{eq:InvAction}
\end{eqnarray}
where the dot (e.g. that in $\XV\cdot \XL$) indicates a certain suitable
operation. If this dot
represents the $G$ transformation $*$
defined by $g\XV*\XL\equiv \delta _G(\XV)\XL$, then
this formula reduces to the original ${\cal H}_{VL}~[={\cal L}(\XV*\XL)]$.  
The $Q$ and $G$ transformation law of ${\cal L}(\XL\cdot \XV)$ may be different
from that of
${\cal H}_{VL}$ only in the terms proportional to $g$. 
For the $Q$ transformation, for instance, we have 
\begin{eqnarray}
\delta _Q(\varepsilon ){\cal L}(\XV\cdot \XL)&=&({\rm total~derivative})\nn
&&+2gi\bar\varepsilon ^i\gamma ^a[W_a,\,\Omega ^j;\,L_{ij}]
+2gi\bar\varepsilon ^i[M,\,\Omega ^j;\,L_{ij}]\nn
 &&+gi\bar\varepsilon \gamma ^a[M,\,W_a;\,\varphi ]
+\myfrac{gi}2\bar\varepsilon \gamma ^{ab}[W_a,\,W_b;\,\varphi ]\nn
&&+gi\bar\varepsilon \gamma ^{abc}[W_a,\,W_b;\,L]\psi _c
+2gi\bar\varepsilon \gamma ^{ab}[M,\,W_a;\,L]\psi _b,
\label{eq:Trf.L}
\end{eqnarray}
where $[A,\,B;\,C]$ denotes the following Jacobi-like operation:
\begin{eqnarray}
 [A,\,B;\,C]&\equiv &A\cdot (B*C)-B\cdot (A*C)-(A*B)\cdot C.\label{eq:tricommutator}
\end{eqnarray}
The $G$ transformation of (\ref{eq:InvAction}) also 
takes a similar form.
Hence if we find a dot operation $(\cdot )$ for which (\ref{eq:tricommutator})
identically vanishes, then the action formula (\ref{eq:InvAction}) 
will be invariant up to the total derivative under the
$Q$ and $G$ transformation, in addition to the $S$ transformation.
(Of course if we choose $\cdot $ as $*$, the operation
(\ref{eq:tricommutator}) vanishes, as the Jacobi identity.)  

For instance, we can see from (\ref{eq:Trf.L}) that
the action formula (\ref{eq:InvAction}) gives an invariant 
by taking the dot operation to be a simple product,
if the vector multiplet is Abelian and the linear multiplet carries no
gauge group charge or is charged only under the Abelian
group of this vector multiplet, like the central charge transformation $
\delta _Z$. When the linear multiplet carries no charges at all, the
constrained vector field $E^a$ can be replaced by the three-form gauge 
field $E_{\mu \nu \lambda }$. Using this, the third line of the above action 
(\ref{eq:InvAction}) can be rewritten, up to total derivative, as  
\begin{equation}
-\myfrac12W_a\left(E^a-2i\bar\psi _b\gamma ^{ba}\varphi +2i\bar\psi _b^{(i}\gamma 
 ^{abc}\psi _c^{j)}L_{ij}\right)\quad \rightarrow \quad 
-\myfrac1{4!}\epsilon ^{\mu \nu \lambda \rho \sigma }F_{\mu \nu }(W)E_{\lambda \rho \sigma }.\label{eq:InvAction2}
\end{equation}
Similarly replacing $*$ by $\cdot $ in
 Eq.~(\ref{eq:Nreplace}), we can obtain another invariant action
 formula written in terms of the four-form gauge field $H_{\mu \nu \rho \sigma }$.
This gives an off-shell formulation of the SUSY-singlet `coupling field' 
introduced in Ref.~\citen{ref:BKP},   
as will be discussed in a forthcoming paper.\cite{ref:KOF}  

 The action formula (\ref{eq:InvAction}) 
can be used to write the action for a general
matter-Yang-Mills system coupled to supergravity.
If we use the above embedding formula, (\ref{eq:VtoL}), of the vector
multiplets into a linear multiplet and apply the action formula
(\ref{eq:InvAction}) and (\ref{eq:InvAction2}),
${\cal L}_V = {\cal L}\left(\XV^A \XL_A(\XV)\right)$, then we obtain a general
Yang-Mills-supergravity action.  Although the action formula can be applied
only to the Abelian vector multiplets $\XV^A$,
interestingly $\XV^A$ can be extended
to include the non-Abelian vector multiplets
$\XV^I$  in this Yang-Mills action; 
that is, the quadratic function $f_A(M)$ multiplied by $M^A$ can 
be extended to a cubic function $\calN(M)$ as
$-\myfrac16f_{A,IJ}M^AM^IM^J\rightarrow \calN=c_{IJK}M^IM^JM^K$.
Also, the action for a general hypermultiplet matter system can be
obtained similarly.  The kinetic term for the hypermultiplet is given 
by ${\cal L}_{H\rm kinetic}={\cal L}\left(\XA~ d^{\alpha \beta } \XL_{\alpha \beta }(\XH,\XZ\XH)\right)$,
with the central charge vector multiplet $\XA=\XV^0$ and an
antisymmetric $G$ invariant tensor $d^{\alpha \beta }$. 
The mass term for the hypermultiplet are given by 
${\cal L}_{H\rm mass}={\cal L}\left(\XA~ \eta ^{\alpha \beta }\XL_{\alpha \beta }(\XH,\XH)\right)$,
with a symmetric $G$ invariant tensor $\eta ^{\alpha \beta }$.
Finally, the action for the unconstrained linear multiplet is given by 
${\cal L}_L={\cal L}\left(\XV(\XL)~ \XL\right)$, which may contain a kinetic term
for the four-form field $H_{\mu \nu \rho \sigma }$ in addition to that for $E_{\mu \nu \lambda }$.
These actions in the superconformal tensor calculus must be identified
with those in the Poincar\'e supergravity tensor calculus. 

In the case that the linear multiplet carries a non-Abelian charge, the
invariant action  cannot be obtained in a simple way:
If we assume that the linear multiplet 
is Lie-algebra valued and interpret the dot operation as a trace, 
then the Jacobi-like operation (\ref{eq:tricommutator}) does not vanish.
 However, it is not impossible to obtain this action. For example,
 one can consider Abelian vector multiplets $\XV^\alpha $ carrying a
non-Abelian charge. That is,
\begin{eqnarray}
{}[ G_I,\,G_\alpha ]&=&f_{I\alpha }{}^\beta G_\beta =-\{\rho (G_I)\}^\beta {}_\alpha G_\beta ,
\qquad [G_\alpha ,\,G_\beta ]=0,
\end{eqnarray}
where $G_I$ and $G_\alpha $
are generators of non-Abelian and Abelian generators,
respectively.
The linear multiplet is also assumed to be an ajoint representation
of this group, $(\XL^I,\XL^\alpha )$.
Then the Jacobi-like operation (\ref{eq:tricommutator}) vanishes 
if we take the dot operation to be given by
\begin{eqnarray}
A\cdot B&=&A_\alpha B^\alpha =\rho _{\alpha \beta }A^\alpha B^\beta =-B\cdot A,\quad \rho _{\alpha \beta }=-\rho _{\beta \alpha }, 
\end{eqnarray} 
and  ${\cal L}(\XV^\alpha \XL^\beta \rho _{\alpha \beta })$ gives an invariant action, while the linear
multiplet carries a non-Abelian charge.
\section{The two-form gauge field and Nishino-Rajpoot formulation
\label{sec:another}}
To this point
in the text, the three-form gauge field $E_{\mu \nu \lambda }$ and 
the four-form gauge fields $H_{\mu \nu \rho \sigma }$ have appeared,
in addition to the one-form gauge fields $W_\mu $.
A two-form gauge field $B_{\mu \nu }$ 
can also be introduced in the process by solving the constraint 
$\XL_0(\XV)=0$, which we impose on a set of Vector multiplets ${\XV^I}$ 
using an embedding quadratic formulation $f_0(M)$. The solution leads to  
another type of the Weyl multiplet
that contains $B_{\mu \nu }$. The formulation
with this new multiplet gives the alternative supergravity presented 
by Nishino and Rajpoot\cite{Nis-Raj} 
after suitable $S$  and $K$ gauge fixing. 
(Thus we will call this the `N-R formulation'.)

Here, we choose the quadratic function $f_0(M)$ to be $G$ inert:
$[A,B]^IC^Jf_{0,IJ}+B^I[A,C]^Jf_{0,IJ}=0$. 
Then, we solve the equation $\XL_0  (\XV)=0$.   
The equation $L^{ij}(\XV)=0$ sets one of the auxiliary fields
$Y^I_{ij}$ of the vector multiplets $\XV^I$ equal to zero.
The equation $\varphi^i(\XV)=0$ makes the auxiliary spinor field $\chi ^i$
of the Weyl multiplet a dependent field, and similarly 
the equation $N(\XV)=0$ is solved with respect to  
the auxiliary scalar field $D$ of the
Weyl multiplet. Then, the equation $E^a(\XV)=0$ becomes a total
derivative in this gauge-invariant case, as mentioned above:
\begin{eqnarray}
E^\mu (\XV)&=&e^{-1}\partial _\nu \left(
\myfrac16\epsilon ^{\mu \nu \lambda \rho \sigma }E_{\lambda \rho \sigma }(\XV)\right)\nn 
&=&e^{-1}\partial _\nu \left\{e\left(4v^{\mu \nu }f+\hat F^{\mu \nu I}(W)f_I
+i\bar\Omega ^I\gamma ^{\mu \nu }\Omega ^Jf_{IJ}\right.\right.\nn
&&\qquad \qquad 
\left.+i\bar\psi _\rho \gamma ^{\mu \nu \rho \sigma }\psi _\sigma f-2i\bar\psi _\lambda \gamma ^{\mu \nu \lambda }\Omega ^If_I
\right)\nn
&&\qquad \qquad \left.+\myfrac12\epsilon ^{\mu \nu \lambda \rho \sigma }\left(W^I_\lambda \partial _\rho W^J_\sigma 
-\myfrac13gW^I_\lambda [W_\rho ,\,W_\sigma ]^J\right)f_{IJ}\right\}\nn
&=&0.\label{eq:eq.B}
\end{eqnarray}
Thus this equation is also solved by making 
the auxiliary tensor field $v_{ab}$ dependent. However, the last
equation here does not fix all components of $v_{ab}$, 
because $E^a$ and $v_{ab}$ have $4$ and $10$ degrees of
freedom, respectively. 
Of course the equation can be solved by means of adding 
the two-form field $B_{\mu \nu }$, which has 6
degrees of freedom:
$0=E_{\mu \nu \lambda }(\XV)+
3\partial _{[\mu }B_{\nu \lambda ]}$.
This can also be rewritten as 
\begin{eqnarray}
0&=&\hat F_{\mu \nu \lambda }(B)
-\myfrac12e \epsilon _{\mu \nu \lambda \rho \sigma }(4v^{\rho \sigma }f+\hat F^{\rho \sigma I}(W)f_I)
+i\bar\Omega  ^I\gamma _{\mu \nu \lambda }\Omega ^Jf_{IJ},
\end{eqnarray}
where $\hat F_{\mu \nu \lambda }(B)$
is the covariant field strength of $B_{\mu \nu }$, which
is given by
\begin{eqnarray}
\hat F_{\mu \nu \lambda }(B)&=&3\partial _{[\mu }B_{\nu \lambda ]}
-6i\bar\psi _{[\mu }\gamma _{\nu \lambda ]}\Omega ^If_I+6i\bar\psi _{[\mu }\gamma _{\nu }\psi _{\lambda ]}f\nn
&&-3\partial _{[\mu }W_\nu ^IW_{\lambda ]}^Jf_{IJ}+W_{[\mu }^I[W_\nu ,\,W_{\lambda ]}]^Jf_{IJ}.
\label{eq:Bstrength}
\end{eqnarray}

A $Q$ and $G$ transformation law of $B_{\mu \nu }$ can be easily found
from the $Q$  and $G$ covariance of (\ref{eq:Bstrength}) in the same
way as that of $H_{\mu \nu \rho \sigma }$ and $E_{\mu \nu \lambda }$ in the linear multiplet 
case. The gauge field $B_{\mu \nu }$
is $S$ invariant and transforms under 
$\delta =\delta _Q(\varepsilon )+\delta _B(\Lambda _\mu ^B)
+\delta _G(\Lambda )$ as
\begin{eqnarray}
\delta B_{\mu \nu }&=&2i\bar\varepsilon \gamma _{\mu \nu }\Omega ^If_I-4i\bar\varepsilon \gamma _{[\mu }\psi _{\nu ]}f
            +(\delta _Q(\varepsilon )W_{[\mu }^I)W_{\nu ]}^Jf_{IJ}\nn
 &&+2\partial _{[\mu }\Lambda ^B_{\nu ]}+\partial _{[\mu }W_{\nu ]}^I\Lambda ^Jf_{IJ}.
\end{eqnarray}
The algebra on $B_{\mu \nu }$, of course, closes, although the algebra is  
modified as follows:
\begin{eqnarray}
[\delta _Q(\varepsilon _1),\,\delta _Q(\varepsilon _2)]&=&({\rm R.H.S.~of~(\ref{eq:QQcommutator2})}) 
+\delta _B\left(-2i\bar\varepsilon _1\gamma _\mu \varepsilon _2f-i\bar\varepsilon _1\varepsilon _2W_\mu ^If_I\right),\nn
{}[\delta _Q(\varepsilon ),\,\delta _G(\Lambda )]&=&\delta _B\left(\delta _Q(\varepsilon )W_\mu ^I\Lambda ^Jf_{IJ}\right),\nn
{}[\delta _G(\Lambda _1),\,\delta _G(\Lambda _2)]&=&
\delta _B\left(-\myfrac12W_\mu ^I[\Lambda _1,\,\Lambda _2]^Jf_{IJ}\right)
+\delta _G\left(-[\Lambda _1,\,\Lambda _2]\right). 
\end{eqnarray}
  Now the constraint equations $\XL_0(\XV)=0$ has replaced the `matter'
sub-multiplet $(v_{ab},\chi ^i,D)$ of the Weyl multiplet 
by the tensor field $B_{\mu \nu }$ and a linear combination 
of the vector multiplets
with one component of the auxiliary field $Y^I_{ij}$ eliminated.
We thus have obtained the N-R formulation with an alternative Weyl
multiplet.
If we take only a single vector multiplet, say the central vector
multiplet, $\XV^0=(\alpha,\,A_\mu,\,\Omega^{0i},\,Y^{0ij})$,
and set $f_0(M)=\alpha ^2$,
then the conventional Weyl multiplet
$(e_\mu{}^a,\,\psi_\mu^i,\,b_\mu,\,V_\mu^{ij},\,v_{ab},\,\chi^i,\,D)$ 
is replaced by a new $32+32$ multiplet consisting of 
$(e_\mu{}^a,\,\psi_\mu^i,\,b_\mu,\,V_\mu^{ij},B_{\mu \nu},
\,A_\mu,\,\alpha,\,\Omega^{0i})$. 

There are two known different types of formulations of 
the on-shell supergravity in 5D, the conventional one with no
$B_{\mu \nu }$ and 
the above N-R formulation. These different
formulations give different physics of course. 
From the point of view of off-shell formulation, this difference is only a
difference in the cubic function $\calN(M)$ that characterizes
super Yang-Mills systems in 5D.         
If there is an Abelian vector multiplet $\XV_{\rm NR}$ that appears
 only in the form
of a Lagrange multiplier, that is, if $\calN$ takes the form  
\begin{eqnarray}
\calN&=&c_{IJK}M^IM^JM^K+M_{\rm NR}f_{IJ}M^IM^J,
\end{eqnarray}
and if matter multiplets carry no charge of this vector multiplet,
then the equations of motion
take the form $\XL_0(\XV)=0$
by the variation of $\XV_{\rm NR}$.
Therefore, after integrating out $\XV_{\rm NR}$,
the N-R formulation appears.   
Conversely, if there is no such Abelian vector multiplet, the
formulation gives the conventional one.      

\section{Summary and comments}
In this paper, we have presented a superconformal tensor calculus in
five dimensions. This work extends a previous work,\cite{Kug-Oha1} which 
presents Poincar\'e supergravity tensor calculus
that is almost completely
derived using dimensional reduction and
decomposition from the known superconformal
tensor calculus in six dimensions. 
The significant difference between the superconformal tensor calculus
presented in this paper
and Poincar\'e one is that the minimal Weyl multiplet in the
superconformal case has 32 Bose plus 32 Fermi degrees of freedom,
while that in the Poincar\'e case
has (40+40) degrees of freedom.  
      
In a previous paper,\cite{Kug-Oha2} we constructed 
off-shell $d=5$ supergravity coupled to a
matter-Yang-Mills system by using the Poincar\'e supergravity
tensor calculus.\cite{Kug-Oha1} There, intricate and 
tedious computations were necessary (owing to a lack of 
$S$ symmetry) to rewrite the Einstein and
Rarita-Schwinger terms into canonical form.
However, now we can write down the same action with 
little work, thanks to the full superconformal symmetry.
Actually, we can show readily that 
this superconformal calculus is equivalent to
two Poincar\'e calculuses with two different $S$ gauge choices,
and thus the two Poincar\'e calculuses are equivalent (see Appendix D).
Also, it is easy to show the equivalence to other Poincar\'e
supergravities.\cite{Gun-Zag,Cer-Dal,Nis-Raj}

There appeared several by-products in the text.
In this calculus, we have not imposed constraints on the $Q$
and $M$ curvatures. 
Though this is a purely technical point and unimportant from the viewpoint
 of physics, it could be interesting to pursue, as it is different
from usual situation in superconformal gravities. 
This formulation with no
constraints makes clear that these constraints are completely
unsubstantial. It is thus seen that
superconformal gravity with various forms of constraints
can describe the same physics.

  Moreover, the four-form gauge field $H_{\mu \nu \rho \sigma }$ and the two-form
gauge field $B_{\mu \nu }$ have appeared in this off-shell formulation in
addition to the three-form gauge field $E_{\mu \nu \lambda }$.
  
In recent studies of the brane world scenario, the four-form gauge
field $H_{\mu \nu \rho \sigma }$ plays an important role in connection with the
$Q$ singlet scalar `coupling field' $G$. Now we can construct an off-shell
formulation of $H_{\mu \nu \rho \sigma }$ and $G$,
 though only an on-shell formulation is known to this time.
 This extension may allow for the extraction of general
properties from the brane world scenario without going into 
the details of the models. 
This will be discussed in a forthcoming paper.\cite{ref:KOF}
Also ${\cal L}\left(\XV(\XL) \XL\right)$ may contain a kinetic 
term for $H_{\mu \nu \rho \sigma }$ and lead to
interesting physics in the brane
world scenario.

Introducing the two-form gauge field $B_{\mu \nu }$ implies a new
non-minimal Weyl multiplet, which should be equivalent 
to that presented by Nishino and Rajpoot.\cite{Nis-Raj} 
From the viewpoint of the off-shell formalism, this system is unified
with the general matter Yang-Mills system.\cite{Gun-Zag,Cer-Dal} 

The tensor multiplet as a matter multiplet, containing a two-form gauge
field $B_{\mu \nu }$, is known in on-shell formulation.
But we have not yet understood it in the present tensor calculus.
Excluding this problem, however, this superconformal tensor calculus
shoud produce all types of supergravity in 5D. 
Superconformal tensor calculus will
provide powerful tools for the brane world scenario from a more unified
viewpoint.
\\ \\

 \section*{Acknowledgements}
The authors would like to thank Professor Taichiro Kugo for helpful
discussions and careful reading of the manuscript.
\appendix
\section{Conventions and Useful Identities}
We employ the notation of Ref.~\citen{Kug-Oha1}.
The gamma matrices $\gam^a$ satisfy
$\{\gam^a,\gam^b\}=2\eta^{ab}$ and $(\gam_a)^{\dagger}=\eta_{ab}\gam^b$,
where $\eta^{ab}={\rm diag}(+,-,-,-,-)$.
\,$\gam_{a\ldots b}$ represents an antisymmetriced product
of gamma matrices:
\beq
\gam_{a\ldots b}=\gam_{[a}\ldots\gam_{b]},
\eeq
where the square brackets denote complete antisymmetrization
with weight 1. Similarly $(\ldots)$ denote complete
symmetrization with weight 1.
We chose the Dirac matrices to satisfy
\beq
\gam^{a_1\ldots a_5}=\eps^{a_1\ldots a_5}
\eeq
where $\eps^{a_1\ldots a_5}$ is a totally antisymmetric
tensor with $\eps^{01234}=1$.
With this choice, the duality relation reads
\beq
\gam^{a_1\ldots a_n}=\frac{(-1)^{n(n-1)/2}}{(5-n)!}
\eps^{a_1\ldots a_nb_1\ldots b_{5-n}}\gam_{b_1\ldots b_{5-n}}.
\eeq

The $SU(2)$ index $i$ ($i$=1,2) is raised and lowered with $\eps_{ij}$,
where $\eps_{12}=\eps^{12}=1$, in the
northwest-southeast (NW-SE) convention:
\beq
A^i=\eps^{ij}A_j,\hspace{2em}A_i=A^j\eps_{ji}.
\eeq
A useful formula in treating these indices is
$A^iB^jC_j=-A^jB_jC^i-A_jB^iC^j$.

The charge conjugation matrix C in 5D has the properties
\beq
C^T=-C,\hspace{2em}C^{\dagger}C=1,\hspace{2em}C\gam_aC^{-1}=\gam_a^T.
\eeq
Our five-dimentional spinors satisfy the $SU(2)$-Majorana condition
\beq
\bpsi^i\equiv\psi_i^{\dagger}\gam^0=\psi^{iT}C,
\eeq
where spinor indices are omitted. When $SU(2)$ indices are suppressed
in bilinear terms of spinors, NW-SE contraction is understood, e.g.
$\bpsi\gam^{a_1\ldots a_n}\lam=\bpsi^i\gam^{a_1\ldots a_n}\lam_i$.
Changing the order of spinors in a bilinear leads to the following
signs:
\beq
\bpsi\gam^{a_1\ldots a_n}\lam=
(-1)^{(n+1)(n+2)/2}\blam\gam^{a_1\ldots a_n}\psi.
\eeq 
If the $SU(2)$ indices are not contracted, the sign becomes opposite.
We often use the Fierz identity, which in 5D reads
\beq
\psi^i\blam^j=-\frac{1}{4}(\blam^j\psi^i)
-\frac{1}{4}(\blam^j\gam^a\psi^i)\gam_a
+\frac{1}{8}(\blam^j\gam^{ab}\psi^i)\gam_{ab}
\eeq
 
\section{Dimensional Reduction to 5D from 6D}
5D conformal supergravity can be obtained
from 6D conformal supergravity\cite{Ber-Sez-Pro}
through dimensional reduction.
Upon reduction to five dimensions,
the 6D Weyl multiplet (40+40) become reducible,
and thus there is a need to decompose this multiplet into
the 5D Weyl multiplet (32+32) and the central charge
vector multiplet (8+8).

Basically, we follow the dimensional reduction procedure
explaned in Ref.~\citen{Kug-Oha1},
to which we refer the reader for the details.
The standard form for the sechsbein $e_{\uM}{}^A$ is
\beq
\unde_{\uM}{}^A=\left(\bear{cc}
\unde_{\mu}{}^a&\unde_{\mu}{}^5\\
\unde_z{}^a&\unde_z{}^5
\eear\right)
=\left(\bear{cc}
e_{\mu}{}^a&\alp^{-1}A_{\mu}\\
0&\alp^{-1}
\eear\right).
\eeq
Here $\uM,\uN,\ldots$ are six dimensional space-time indices
and, $z$ denotes fifth spatial direction
wheras $A,B,\ldots$ denote six-dimensional local Lorentz indices.
The underlined fields are the components of the
six-dimensional Weyl or matter multiplet.
\,$\alp$ and $A_{\mu}$ are identified with the scalar and the vector
components of the central charge vector multiplet.
The relation between tensors in 6D and 5D is given by the following rule:
{\it Tensors with flat indices only are the same in 6D and 5D}.
Thus, for a vector, for example, we have $\underline{v}_a=v_a$
(so we need not use an underbar for flat indices),
but
\beq
\underline{v}_{\mu}=\unde_{\mu}{}^av_a+\unde_{\mu}{}^5v_5
=v_{\mu}+A_{\mu}\underline{v}_z.
\eeq
We decompose the six-dimensional gamma matrices $\Gam^M$ and
the charge conjugation matrix $\uC$ as
\beqa
\Gam^a\eqn{=}\gam^a\otimes\sigma_1,
\quad
\Gam^5=1\otimes i\sigma_2,\CR
\uC\eqn{=}C\otimes i\sigma_2.
\eeqa
The six-dimensional chirality operator
$\Gam_7$ is written $\Gam_7=1\otimes\sigma_3$.
The six dimensional $SU(2)$-Majorana-Weyl spinor $\psi_{\pm}^i$,
which satifies the $SU(2)$-Majorana condition
$\bpsi_{\pm}^i\equiv\psi_{i\pm}^{\dagger}\Gam^0=\psi_{\pm}^{iT}\uC$
and the Weyl condition $\Gam_7\psi_{\pm}^i=\pm\psi_{\pm}^i$,
is decomposed as
\beq
\psi^i_+=\psi^i\otimes\left(\bear{c}
1\\
0
\eear\right),\hspace{2em}
\psi^i_-=
i\psi^i\otimes
\left(\bear{c}
0\\
1
\eear\right),
\eeq
where $\psi^i$ is a five-dimensional $SU(2)$-Majorana spinor.

The generators of 6D superconformal algebra,
which is Osp$(8^*|2)$, are labeled
\beq
P_A,\:\:M_{AB},\:\:K_A,\:\:D,\:\:U_{ij},
\:\:Q^i_{\alp},\:\:S^i_{\alp}.
\eeq
Of these, the generators $M_{a5}$ and $K_5$ are redundant in 5D
and are used to fix redundant gauge fields, as described below.
The independent gauge fields of the $N=2, d=6$ Weyl multiplet,
which realize this algebra are
\beq
\unde_{\uM}{}^A,\:\:\undpsi_{\uM +}^i,\:\:\undb_{\uM},\:\:\uV_{\uM}^{ij},
\:\:\uT^-_{ABC},\:\:\undchi^i_-,\:\:\uD.
\eeq
The first four are the gauge fields corresponding to
the generators $P_A$,\,$Q_{\alp}^i$,\,$D$ and $U^{ij}$.
The last three are the additive matter fields.
The other gauge fields, $\undome_{\uM}{}^{AB}$, $\undphi_{\uM}^i$
and $\undf_{\uM}{}^A$, which correspond to the generators $M_{AB}$,
$S^i_{\alp}$ and $K_A$ can be expressed in terms of the above
independent
gauge fields by imposing the curvatures constraints
\beqa
&&\uhR_{\uM\uN}{}^A(P)=0,\CR
&&\uhR_{\uM\uN}{}^{AB}(M)\unde^{\uN}{}_B+\uT^-_{\uM BC}\uT^{-ABC}
+\frac{1}{12}\unde_{\uM}{}^A\uD=0,\CR
&&\Gam^{\uN}\uhR_{\uM\uN}^i(Q)=-\frac{1}{12}\Gam_{\uM}\undchi^i,\label{b10}
\eeqa
where $\uhR_{\uM\uN}{}^A(P)$, $\uhR_{\uM\uN}{}^{AB}(M)$
and $\uhR_{\uM\uN}^i(Q)$ are the $P$, $M$ and $Q$ curvatures,
respectively.
(For more details, see Ref. \citen{Ber-Sez-Pro}, but
note that we use the notation of Ref. \citen{Kug-Oha1}.)

First, we decompose the six-dimensional Weyl multiplet
into three classes,
\beqa
&&(\,\unde_{\mu}{}^a,\,\undpsi_{\mu}^i,\,\undb_{\mu},\,\uV_{\mu}^{ij},
\,\uT^-_{ab5},
\,\undchi^i,\,\uD\,),\CR
&&(\,\unde_{\mu}{}^5,\,\unde_{z}{}^5,\,\undpsi_z^i,\,\uV_z^{ij}\,),\CR
&&(\,\unde_z{}^a,\,\undb_z\,).
\eeqa
Roughly speaking, the first class gives the five-dimensional
Weyl multiplet and the second the central charge
vector multiplet.
The last class consists of redundant gauge fields.
Redundant gauge fields can be set equal to zero as a
gauge-fixing choice for the
redundant $M_{a5}$ and $K_5$ symmetries.
However, the condition $\unde_z{}^a=\undb_z=0$ is not invariant
under $Q$ and $S$ transformations.
Thus, we have to add 
a suitable gauge transformation to the original
$Q$ and $S$ transformations.
Explicitly, the original $Q$ transformation of $\unde_z{}^a$
is
\beq
\delta_Q^{\rm 6D}(\veps)\unde_z{}^a=-2i\bveps\gam^a\undpsi_z.
\eeq
Adding a $M_{a5}$ transformation with parameter
$\theta_M^{a5}=-2i\alp\bveps\gam^a\undpsi_z$
to the original
$Q$ transformation, we
obtain a $Q$ transformation, under which the constraint
$\unde_z{}^a=0$ remains invariant.
Similarly, in order to keep $\undb_z=0$ invariant 
under $Q$ and $S$ transformations,
we should add a $K_5$ transformation with
parameter $\theta_K^5(\veps)=i\bveps\undchi/24
+i\bveps\undphi_5$ to the original $Q$ transformation
and a $K_5$ transformation
with parameter $\zeta_K^5(\eta)=i\bareta\undpsi_z$
to the original $S$ transformation, $\delta_S^{\rm 6D}(\eta)$.

In the original gauge transformation law for fields of the first class,
the central charge vector multiplet components do not decouple.
For example, the $Q$ transformation of $\undpsi_{\mu}^i$ is
\beq
\delta_Q^{\rm 6D}(\veps)\undpsi_{\mu}^i=
\cD_{\mu}\veps^i-\frac{1}{4}\uT^{\rs}{}_5^-\gam_{\mu\rs}\veps^i
+\frac{1}{2\alp}\partial_{[\mu}A_{\nu]}\gam^{\nu}\veps^i
-2i(\bveps\gam_{\mu}\undpsi_z)\undpsi^i_z+\ldots\ .
\eeq
This transformation includes $\alp=\unde^z{}_5$,
$A_{\mu}=\alp\unde_{\mu}{}^5$ and
$\undpsi_z$, which are the fields of the central charge vector multiplet.
To get rid of these fields from the transformation law of
the five-dimensional Weyl multiplet,
we need to redefine the gauge fields and the $Q$ and
$S$ transformations.
The proper identification of the central charge vector multiplet
components turns out to be
\beq
\alp=\unde^{z}{}_5,\quad A_{\mu}=\alp \unde_{\mu}{}^5,
\quad\Ome_0^i=-\alp^2\undpsi_z^i,
\quad Y_0^{ij}=\alp^2\uV_z^{ij}-\frac{3i}{\alp}\bOme_0^i\Ome_0^j.
\eeq
The field $\alp$, whose Weyl weight is 1, is used to adjust
the Weyl weight of the
redefined field.
For example, $A_{\mu}$ should carry Weyl weight $0$
as any gauge field,
but the Weyl weight of $\unde_{\mu}{}^5$ is $-1$,
so we identify $\alp\unde_{\mu}{}^5$ with the gauge field $A_{\mu}$.
The correction term $3i\bOme_0^i\Ome_0^j/\alp$ in
the redefinition of $Y_0^{ij}$ is needed to
remove the central charge vector multiplet
from the algebra.

Similarly, the irreducible (32+32) Weyl multiplet in 5D is
identified as
\beqa
e_{\mu}{}^a\eqn{=}\unde_{\mu}{}^a,\quad\psi_{\mu}^i=\undpsi_{\mu}^i,
\quad b_{\mu}=\undb_{\mu},\CR
V_{\mu}^{ij}\eqn{=}\uV_{\mu}^{ij}+\frac{2i}{\alp}
\bpsi_{\mu}^{(i}\Ome_0^{j)}
-\frac{i}{\alp^2}\bOme_0^i\gam_{\mu}\Ome_0^j,\CR
v_{ab}\eqn{=}-\uT_{ab5}^--\frac{1}{4\alp}\hF_{ab}(A)
+\frac{i}{2\alp^2}\bOme_0\gam_{ab}\Ome_0,\CR
\chi^i\eqn{=}\frac{16}{15}\undchi^i
+\frac{8}{5\alp}\left(\nchD\Ome_0^i+\frac{1}{2\alp}(\nchD\alp)\Ome_0^i
+\frac{3}{2}\gam\cdot v\Ome_0^i\right)\CR
&&-\frac{1}{5}\left(\gam\cdot\hR^i(Q)-\frac{6}{\alp^2}\gam\cdot\hF(A)\Ome_0^i
\right)
-\frac{8}{\alp^2}t^{ij}\Ome_{0j}
-\frac{2i}{\alp^3}\gam_{ab}\Ome_0^i\bOme_0\gam^{ab}\Ome_0,\CR
D\eqn{=}\frac{8}{15}\uD-\frac{1}{10}\hR_{ab}{}^{ab}(M)
+2v_{ab}v^{ab}\CR
&&+\frac{1}{\alp}\left(i\bOme_0\chi+\frac{4i}{5}\bOme_0\gam\cdot\hR(Q)\right)
-\frac{4}{5\alp}\chD_a\chD^a\alp
-\frac{2}{5\alp^2}\chD^a\alp\chD_a\alp\CR
&&+\frac{2}{5\alp^2}\hF_{ab}(A)\hF^{ab}(A)
-\frac{4}{\alp^2}t_{ij}t^{ij}
+\cO(\Ome_0^2),
\eeqa
where $\cO(\Ome_0^2)$ represents terms of higher order in $\Ome_0$.

The relation between the $Q$ and $S$ transformations in 5D and in 6D
is finally given by
\beqa
\delta_Q(\veps)\eqn{=}\delta_Q^{\rm 6D}(\veps)
+\delta_M(\theta_M(\veps))+\delta_U(\theta_U(\veps))
+\delta_S(\theta_S(\veps))+\delta_K(\theta_K(\veps)),\CR
\delta_S(\eta)\eqn{=}\delta_S^{\rm 6D}(\eta)
+\delta_K(\zeta_K(\eta)),
\eeqa
where
\beqa
&&\theta_M^{a5}(\veps)=\frac{2i}{\alp}\bveps\gam^a\Ome_0,\quad
\theta_U^{ij}(\veps)=-\frac{2i}{\alp}\bveps^{(i}\Ome_0^{j)},\CR
&&\theta_S^i(\veps)=\frac{1}{4}\gam
\cdot\left(-v+\frac{1}{4\alp}\hF(A)\right)\veps^i
-\frac{i}{2\alp^2}(\bOme_0^i\Ome_0^j)\veps_j
+\frac{i}{2\alp^2}(\bOme_0^i\gam_a\Ome_0^j)\gam^a\veps_j,\CR
&&\theta_{K\mu}(\veps)=-\frac{i}{24}\bveps\gam_{\mu}\undchi
-i\bveps\left(\undphi_{\mu}-\theta_{S\mu}(\frac{\Ome_0}{\alp})
-\phi_{\mu}\right)
-i\bveps\phi_a(Q),\CR
&&\theta_K^5(\veps)=\frac{i}{24}\bveps\undchi
+i\bveps\left(\undphi_5-\theta_S(\frac{\Ome_0}{\alp})\right),\quad
\zeta_K^5(\eta)=-\frac{i}{\alp}\bareta\Ome_0.
\eeqa
Here, the dependent gauge fields $\undphi^i_{\mu}$
and $\undphi_5^i$ are those
determined by the curvature constraint (\ref{b10}).

The 6D vector multiplet consists of
a real vector field $\uW_M$, an $SU(2)$-Majorana spinor $\uOme^i$,
and a triplet of the auxiliary scalar field $\uY^{ij}$,
whereas the 5D vector multiplet consists of
a real vector $M_{\mu}$, a scalar $M$, an $SU(2)$ Majorana spinor
$\Ome^i$, and a triplet of the auxiliary scalar $Y^{ij}$.
The proper identification of the vector multiplet
components is
\beqa
&&M=-\uW_5,\quad W_{\mu}=\uW_{\mu},\quad
\Ome^i=\uOme^i{}+\frac{M}{\alp}\Ome_0^i,\CR
&&Y^{ij}=\uY^{ij}+\frac{M}{\alp}Y_0^{ij}
-\frac{2i}{\alp}\bOme_0^{(i}\left(\Ome^{j)}
-\frac{M}{\alp}\Ome_0^{j)}\right).
\eeqa

The 6D linear multiplet consists of
a triplet $\uL^{ij}$, an $SU(2)$-Majorana spinor $\undvphi^i$,
and a constrained vector field $\uE_A$.
The components of the 5D linear multiplet
are identified as
\beqa
&&L^{ij}=\frac{1}{\alp}\uL^{ij},\quad
\vphi^i=
\frac{1}{\alp}(\undvphi^i-2\Ome_{0j}L^{ij}),\CR
&&E^a=\frac{1}{\alp}\left(\uE^a+\frac{4i}{\alp}
\bOme_0^i\gam^a\Ome_0^jL_{ij}-2i\bOme_0\gam^a\vphi\right),\CR
&&N=-\frac{1}{\alp}(\uE_5+2L^{ij}t_{ij}+4i\bOme_0\vphi).
\eeqa

The 6D nonlinear multiplet consists of
a scalar $\uPhi_{\alp}^i$, an $SU(2)$-Majorana spinor
$\undlam^i$, and a vector field $V_A$.
Identification of the 5D nonlinear multiplet is
given by
\beqa
&&\Phi_{\alp}^i=\uPhi_{\alp}^i,\quad
\lam^i=\undlam^i-\frac{1}{\alp}\Ome_0^i,\CR
&&V_a=\uV_a+\frac{1}{\alp}\cD_a\alp,\quad
V^5=-\uV_5-\frac{2i}{\alp}\bOme_0\lam.
\eeqa

The 6D hypermultiplet consists of
a scalar ${\cal \uA}_{\alp}^i$ and a $SU(2)$-Majorana spinor
$\undzeta^{\alp}$.
The 5D hyper multiplet is identified as
\beq
\calA_{\alp}^i=\frac{1}{\sqrt{\alp}}{\cal\uA}_{\alp}^i,\quad
\zeta_{\alp}=\frac{1}{\sqrt{\alp}}
\undzeta_{\alp}+\frac{1}{\alp}\Ome_0^j\calA_{j\alp}.
\eeq

\section{Weyl Multiplet in 4D with no
Constraints\label{sec:no-constr4D}}
In the text we have explained that the constraint for $Q$ and
$M$ curvatures is not necessarily needed.
This fact is not familiar, so we illustrate our formulation
by taking an example of the
well-known $N=1, d=4$ superconformal Weyl multiplet.\cite{n1d41}

The independent gauge fields of the $N=1, d=4$ Weyl multiplet
are the vierbein $e_{\mu}{}^a$, the gravitino $\psi_{\mu}$,
the $D$ gauge field $b_{\mu}$, and the $U(1)$ gauge field $A_{\mu}$.
In this section, $\mu,\nu,\ldots$ and $a,b,\ldots$
are four-dimensional indices.
The spinors are Majorana.
In the usual formulation, in which the $Q$ curvature constraint
$\gam^{\nu}\hR_{\mn}(Q)=0$ is imposed,
the
$Q$,\,$S$ and $K$ transformation laws of the Weyl multiplet
are given by
\beqa
\delta e_{\mu}{}^a\eqn{=}-2i\bveps\gam^a\psi_{\mu},\CR
\delta\psi_{\mu}\eqn{=}\cD_{\mu}\veps+i\gam_{\mu}\eta,\CR
\delta b_{\mu}\eqn{=}-2\bveps\phi_{\mu}^{\rm sol}
+2\bareta\psi_{\mu}-2\xi_{k\mu},\CR
\delta A_{\mu}\eqn{=}-4i\bveps\gam_5\phi_{\mu}^{\rm sol}
+4i\bareta\gam_5\psi_{\mu},\label{C1}
\eeqa
where $\phi_{\mu}^{\rm sol}$ is the solution of $\gam^{\nu}\hR_{\mn}(Q)=0$.
We note that
$\hR(Q)$ contains the $S$ gauge field $\phi_{\mu}$
in the form
$\hR_{\mn}(Q)=\hOme_{\mn}(Q)
-2i\gam_{\mu}\phi_{\nu}$,
$\hOme_{\mn}(Q)\equiv\hR_{\mn}(Q)|_{\phi=0}$. 
Solving this with respect to $\phi_{\mu}$,
we have
\beqa
\phi_{\mu}\eqn{=}\phi_{\mu}(Q)+\phi_{\mu}^{\rm sol},\CR
\phi_{\mu}(Q)\eqn{\equiv}\frac{i}{3}\gam^{\nu}\hR_{\mn}(Q)-\frac{i}{12}
\gam_{\mu}{}^{\nu\rho}\hR_{\nu\rho}(Q),\CR
\phi_{\mu}^{\rm sol}\eqn{\equiv}-\frac{i}{3}\gam^{\nu}\hOme_{\mn}
(Q)+\frac{i}{12}
\gam_{\mu}{}^{\nu\rho}\hOme_{\nu\rho}(Q).
\eeqa
Under the $Q$ curvature constraint $\gam^{\nu}\hR_{\mn}(Q)=0$,
\,$\phi_{\mu}$ equals $\phi_{\mu}^{\rm sol}$.
Then we can replace $\phi_{\mu}^{\rm sol}$
in (\ref{C1})
formally by $\phi_{\mu}-\phi_{\mu}(Q)$,
because $\phi_{\mu}(Q)=0$,
and obtain
\beqa
\delta b_{\mu}\eqn{=}-2\bveps\phi_{\mu}+2\bveps\phi_{\mu}(Q)
+2\bareta\psi_{\mu}-2\xi_{k\mu},\CR
\delta A_{\mu}\eqn{=}-4i\bveps\gam_5\phi_{\mu}
+4i\bveps\gam_5\phi_{\mu}(Q)
+4i\bareta\gam_5\psi_{\mu}.
\eeqa
In these expressions,
$\phi_{\mu}$ is decoupled from the transformation laws,
since $\phi_{\mu}$ is cancelled by that in $\phi_{\mu}(Q)$.
This means that the $Q$ curvature constraint is not needed.
We arrive at the following conclusion.
In order to move to the formulation
where the $Q$ and $M$ curvature constraint
is not imposed,
we only have to replace $\phi_{\mu}^{\rm sol}$ by
$\phi_{\mu}-\phi_{\mu}(Q)$.
\section{Equivalency to Poincar\'e Supergravities\label{sec:relation}}
In a previous paper,\cite{Kug-Oha2} which we refer to as II henceforth,
supergravity coupled to a matter-Yang-Millos system in 5D is derived  
on the basis of the supergravity tensor calculus 
presented in Ref. \citen{Kug-Oha1}
which we refer to as I henceforth. However, there, a quite laborious
computation was required to obtain the canonical
form of the Einstein and Rarita-Schwinger terms. 
This is due to the redefinitions of fields, 
in particular that of the Rarita-Schwinger field $\psi _\mu ^i$, (5$\cdot$3)
in II. 
These redefinitions are also accompanied by
modification of the transformation laws, (6$\cdot$8)-(6$\cdot$10) in II.  
Since we have a full superconformal tensor calculus, it is now easy to 
reproduce the Poincar\'e tensor calculus constructed in I and II.
The point is that we can obtain this calculus simply by 
fixing the extraneous gauge freedoms, without making
laborious redefinitions of the gauge field. 
\subsection{Paper I}
First, we identify one of the Abelian vector multiplets with the
central charge vector multiplet, which is a sub-multiplet of the Weyl
multiplet in the Poincar\'e supergravity formulation:
\begin{eqnarray}
(W_\mu ,\,M,\,\Omega ^i,Y^{ij})&=&(A_\mu ,\,\alpha ,\,\Omega ^i=0,\,-t^{ij}\alpha ),
\end{eqnarray}
where the scalar $\alpha $ is covariantly constant.
That is, we choose the following $S$  and $K$ gauges:
\begin{equation}
\XS: \Omega ^i=0,\qquad \XK:\alpha ^{-1}\hatD_a\alpha =0.\label{eq:SKgauge1}
\end{equation}
These gauge fixings are achieved by redefinitions of 
the $Q$ transformation:
\begin{eqnarray}
\tilde \delta _Q(\varepsilon )&=&\delta _Q(\varepsilon )+\delta _S(\eta ^i(\varepsilon ))+\delta _K(\xi _K^a(\varepsilon )),\nn
\eta (\varepsilon )^i&=&-\myfrac1{4\alpha }\gamma \dt \hat F(A)\varepsilon ^i
-t^i{}_j\varepsilon ^j,\nn
\xi _K^a(\varepsilon )&=&i\bar\varepsilon \left(\phi ^a-\phi ^a(Q)\right)+\bar\eta (\varepsilon )\psi ^a\nn
&=&i\bar\varepsilon \left(\phi ^a-\eta (\psi ^a)-\phi ^a(Q)\right).
\end{eqnarray}
Actually, the gauge choices (\ref{eq:SKgauge1}) are invariant under
$\tilde \delta _Q(\varepsilon )$. 
Next, we replace the auxiliary fields of the Weyl multiplet, the vector
multiplet and the linear multiplet as follows: 
\begin{eqnarray}
v_{ab}&\rightarrow &v_{ab}-\myfrac1{2\alpha }\hat F_{ab}(A),
\qquad \chi ^i\rightarrow 16\tilde \chi ^i-\gamma \dt \hat{\calR^i}(Q),\nn
D&\rightarrow &8C+\myfrac12\hat{\calR}(M)-6v^2+\myfrac2\alpha v\dt \hat F(A)
+\myfrac3{2\alpha ^2}\hat F(A)^2+20t^i{}_jt^j{}_i,\nn
Y^{ij}&\rightarrow &Y^{ij}-Mt^{ij},\qquad 
N\rightarrow N+2t^{ij}L_{ij}.        
\end{eqnarray}
Then, we can show that the Poincar\'e supergravity tensor
calculus in the paper I is exactly reproduced.
These gauge choices and the redefinition of the $Q$ transformation must be 
accompanied by redefinitions of the full covariant curvature $\hatR_
{\mu \nu }{}^A$ and the covariant derivative $\hatD_\mu $.
However, such redefinitions are carried out automatically, and there is no
need to do so by hand. 

\subsection{Paper II}
The conditions (5$\cdot$1) and (5$\cdot$6) in II require the gauge fixings
\begin{equation}
\XD: {\calN}=1,\qquad \XS: \Omega ^{Ii}\calN_I=0,\qquad 
\XK: {\calN}^{-1}\hatD_a{\calN}=0.
\end{equation}
These gauge fixings are achieved by
\begin{eqnarray}
\tilde \delta _Q(\varepsilon )&=&\delta _Q(\varepsilon )+\delta _S(\eta ^i(\varepsilon ))+\delta _K(\xi _K^a(\varepsilon )),\nn
\eta (\varepsilon )^i&=&-\myfrac{\calN_I}{12\calN}\gamma \dt \hat F^I(W)\varepsilon ^i
+\myfrac{\calN_I}{3\calN}Y^{Ii}{}_j\varepsilon ^j
+\myfrac{\calN_{IJ}}{3\calN}\Omega ^{Ii}2i\bar\varepsilon \Omega ^J\nn
&=&-\myfrac13\left(\pr{\bf \Gamma }-\gamma \dt v\right)\varepsilon ^i,\nn
\xi _K^a(\varepsilon )&=&i\bar\varepsilon \left(\phi ^a-\eta (\psi ^a)-\phi ^a(Q)\right),
\end{eqnarray}
and the following replacements are needed:
\begin{eqnarray}
\myfrac{\calN_I}{3\calN}Y^{Iij}&\rightarrow &-\tilde t^{ij},
\qquad V_\mu ^{ij}\rightarrow \tilde V_\mu ^{ij},\qquad v_{ab}\rightarrow \tilde v_{ab},\nn
\chi ^i&\rightarrow &16\tilde \chi ^i+3\gamma \dt \hat{\calR}^i(Q),\qquad 
D\rightarrow 8\tilde C-\myfrac32\hat{\calR}(M)+2v^2,\nn
\Omega &\rightarrow &\lambda ,\qquad \zeta _\alpha \rightarrow \xi _\alpha ,\qquad 
Y^{Iij}\rightarrow \tilde Y^{Iij}-M^I\tilde t^{ij}.
\end{eqnarray}
The resultant $Q$ transformation laws of the Weyl multiplet, 
the vector multiplet 
and the hypermultiplet are equivalent to (6$\cdot$8), (6$\cdot$9)
and (6$\cdot$10) in II ,respectively. 

Note added: After finishing the original form of this paper,
the authors became aware of the preprint by Bergshoeff et al.
\cite{ref:BCDWHP}
treating conformal supergravity in five dimensions.
They discussed the two versions of the superconformal Weyl multiplets,
which they call the Standard one and the Dilaton one. These correspond
to the conventional one and N-R one in the present paper.
They did not present the full superconformal tensor calculus
which we have given here.


\begin{thebibliography}{99}

\bibitem{Cre}
E.~Cremmer,
{\it Superspace and supergravity}, ed. S.~Hawking
and M.~M.~Ro\v{c}ek (Cambridge University Press, 1980).
%

A.~H.~Chamseddine and H.~Nicolai,
Phys. Lett. {\bf B96} (1980), 89.
%

M.~G\"{u}naydin, G.~Sierra and P.~K.~Townsend,
Nucl. Phys. {\bf B242} (1984), 244;
%
%
%
Nucl. Phys. {\bf B253} (1985), 573.

%
\bibitem{Gun-Zag}
M.~G\"{u}naydin and M.~Zagermann,
Nucl. Phys. {\bf B572} (2000), 131, hep-th/9912027.

\bibitem{Cer-Dal}
Nucl. Phys. {\bf B585} (2000), 143, hep-th/0004111.

\bibitem{Nis-Raj}
H.~Nishino and S.~Rajpoot,
hep-th/0011066.

\bibitem{Ber-Her}
K. Behrndt, C. Herrmann, J. Louis and S. Thomas, 
hep-th/0008112.
%
\bibitem{Mal}
J.~Maldacena,
Adv. Theor. Math. Phys. {\bf 2} (1998), 231, hep-th/9711200.
%
%

%
%
\bibitem{RS1}
L.~Randall and R.~Sundrum,
Phys. Rev. Lett. {\bf 83} (1999), 3370, hep-ph/9905221.

\bibitem{RS2}
L.~Randall and R.~Sundrum,
Phys. Rev. Lett. {\bf 83} (1999), 4690, hep-th/9906064.

%
 
\bibitem{Kug-Oha1}
T.~Kugo and K.~Ohashi,
Prog. Theor. Phys. {\bf 104} (2000), 835, hep-ph/0006231.

\bibitem{Kug-Oha2} 
T.~Kugo and K.~Ohashi,
\PTP{105,2001,323}, hep-ph/0010288.

\bibitem{Zuk12}
 M.~Zucker,
 \NP{B570,2000,267}, hep-th/9907082; 
J. High Energy Phys. 08 (2000), 016, hep-th/9909144.

\bibitem{Zuk3}
M.~Zucker, 
hep-th/0009083.


\bibitem{Nah}
W.~Nahm,
Nucl. Phys. {\bf 135} (1978), 149.

A.~Van Proeyen,
Annals of the University of Craiova, Physics AUC, Volume 9 (part I) 1999,
hep-th/9910030.

\bibitem{n1d41}
M.~Kaku, P.~K.~Townsend and P.~Van Nieuwenhuizen,
Phys. Rev. Lett. {\bf 39} (1977), 1109;
%
%
%
Phys. Lett. {\bf B69} (1977), 304;
%
%
%
Phys. Rev. {\bf D17} (1978), 3179.

\bibitem{n2d41}
B.~de Wit and J.~W.~Van Holten,
Nucl. Phys. {\bf B155} (1979), 530.
%

B.~de Wit, J.~W.~Van Holten and A.~Van Proeyen,
Nucl. Phys. {\bf B167} (1980), 186; {\bf B172} (1980), 543 (E).
%

M.~de Roo, J.~W.~Van Holten, B.~de Wit and A.~Van Proeyen.
Nucl. Phys. {\bf B173} (1980), 175.

\bibitem{n4d4}
E.~Bergshoeff, M.~de Roo and B.~de Wit,
Nucl. Phys. {\bf B192} (1981), 173.

\bibitem{Ber-Sez-Pro}
E.~Bergshoeff, E.~Sezgin and A.~Van Proeyen,
Nucl. Phys. {\bf B264} (1986), 653.

\bibitem{n4d6}
E.~Bergshoeff, E.~Sezgin and A.~Van Proeyen,
Class. Quant. Grav. {\bf 16} (1999), 3193, hep-th/9904085.

%
\bibitem{Ber-deR-deW}
E.~Bergshoeff, M.~de Roo and B.~de Wit,
Nucl. Phys. {\bf B217} (1983), 489.
%

\bibitem{ref:BKP}
E.~Bergshoeff, R.~Kallosh and A.~Van Proeyen, 
J. High Energy Phys. 10 (2000), 033, hep-th/0007044.

\bibitem{ref:KOF}
T.~Fujita, T.~Kugo and K.~Ohashi, in preparation.

\bibitem{ref:BCDWHP}
E.\,Bergshoeff, S.\,Cucu, M.\,Derix, T.\,de Wit, R.\,Halbersma and 
A.\,Van\,Proeyen, 
hep-th/0104113.
   
\end{thebibliography}
\end{document}